
\def\hb{\hfil\break}
\magnification=1000
\overfullrule=0pt
\nopagenumbers
\parindent=10truept
\hsize=12truecm
\vsize=18truecm
\baselineskip=12truept

\noindent
{\bf Pairing on Small Clusters in the Peierls-Hubbard Model: \hb
Implications for C$_{60}$}
\vskip 12truept \noindent
J.~Tinka Gammel$^a$
F.~Guo$^b$, D.~Guo$^b$, K.C.~Ung$^b$, and S.~Mazumdar$^b$,
\vskip 12truept \noindent
{$^a$Materials Research Branch,
NRaD, San Diego, CA 92152-5000, USA}\hb
{$^b$Dept.~of Physics, University of Arizona, Tucson, AZ 85721, USA}
\vskip 2.5 truecm \noindent
{\bf Abstract.}
We study the pairing within the Peierls-Hubbard Model for
electron- and hole-doped analogs of C$_{60}$ accessible to exact
diagonalization techniques (cube, truncated tetrahedron, {\it etc.}).
We discuss how inclusion of electron-phonon interactions can
substantially modify the conclusions about pairing obtained when this
coupling is neglected.  We also discuss potential pitfalls in the
extrapolation from these small system calculations to C$_{60}$, and
stress the necessity of having the correct intuitive picture.
\vskip 12truept \noindent
{\bf 1. Introduction}
\vskip 12truept \noindent
A great deal of numerical effort has gone into investigating
whether pairing of the Cooper type occurs in the weakly doped
two-dimensional Mott-Hubbard insulator. More recently, such
a pairing mechanism has been proposed [1] for doped $C_{60}$.
Specifically, these calculations involve determination of the ground
state energies of the half-filled band system, and the electronic
energies needed to add one and two electrons (holes). Pairing is
supposed to occur if the quantity
$$\Delta E=2E(N\pm 1)-E(N)-E(N\pm 2) \eqno(1)$$ \noindent
is positive.
As approximate analytic techniques have been known to lead to
qualitatively incorrect predictions for
solid state systems with strong electron correlation,
exact numerical techniques --
exact calculation for a (small) finite clusters
followed by extrapolation to the system size of interest --
have been employed [2,3].
Within the Hubbard model,
$$
H_{Hub}= \sum_{\langle ij\rangle,\sigma} t_{ij} c_{i,\sigma}^+c_{j,\sigma}
+U\sum_in_{i,\uparrow}n_{i,\downarrow}~,
\eqno(2) $$
\noindent
where $\langle ij \rangle$ are nearest neighbors and $t_{ij}$=$-$$t_0$,
pairing is found for the cube and the truncated tetrahedron [2,3,4].
\vfil\eject
\noindent {\bf 2. Geometry, Distortion, and Pairing}
\vskip 12truept \noindent
For the geometries shown in Fig.~1, we have calculated
the pair binding energy given in Eq.~1. The result is shown
in Fig.~2.  
We see that there appear to be two regions where pairing can occur:
one at small $U$ and one near $U/t_0$=10.
Our investigation indicates this oscillation may be related to
(avoided) energy level crossings.
Interestingly, large $U$ pairing is found in
the pentagonal prism, indicating
that it is not limited  to bipartite lattices.
The pairing at small $U$ appears to be related to the presence
of energy level degeneracies at $U$=0.

\topinsert
\vglue 3.1truecm
\noindent Fig.~1.
Geometries for which the pair binding energy
has been calculated: (a) drum, $N$=8,10,12;
(b) twisted drum, $N$=8,10,12;
(c) truncated tetrahedron.
\vglue 12.4truecm
\noindent Fig.~2.
Pairing in undistorted clusters
for electrons (solid) and holes (dashed):
(a,c,e) drums, (b,d,f)
twisted drums, (g) truncated tetrahedron, (a,b) $N$=8,
(c,d) $N$=10, (e,f,g) $N$=12.
Note there are two regions, small and large $U$,
where pairing can occur.
\endinsert

In Ref.~{4} we discussed how the variation in
$U$-dependence of the Jahn-Teller distortion with filling masks this
Jahn-Teller driven pairing (present on finite chains only!)
as an electron-electron driven pairing at small $U$.
Basically,
by not letting the system distort, we are overestimating the energies
in each case, but this overestimation is much
stronger for the odd electron case than for the even electron case.
The calculated energy required
to add one hole (or electron) to the undistorted system is therefore
too high, giving ``pairing''.

 To investigate the effects of distortion numerically, we first
add a fixed external distortion $\delta_{ij}$ from uniform
bond length $d_0$ and use
$t_{ij}$=$-t_0+\alpha\delta_{ij}$ in the Hamiltonian Eq.~2.
In Fig.~3 we show the results of our calculation for the truncated
tetrahedron.  For the undistorted case (Fig.~3), we merely reproduced
the results of White {\it et al.} [2] --- pairing was seen to
occur at small $U$. With short
bonds connecting the triangles -- the distortion preferred
at 1/2-filling -- the pairing still exists. This is however not the
Jahn-Teller mode, the noninteracting
limit still having degeneracies. The pairing at small $U$ disappears
for the Jahn-Teller distorted truncated tetrahedron (calculated as the
self-consistent distortion for $N_{el}$=$N$$\pm$1, see below). This
clearly indicates
the relationship between pairing at small $U$ and the
degeneracies in the single particle limit:
in the absence of distortion,
the energy to add one hole or electron is being overestimated.
In Ref.~{4} we showed for the cube that the pairing at large $U$ survives
the $U$=0 Jahn-Teller distortion. However, since there
appears to be an energy-level crossing
it may be that a different geometry than the $U$=0 one should
be considered the ``large-$U$ Jahn-Teller" mode,
and we are investigating this further.

\topinsert
\vglue 5.9truecm
\noindent Fig.~3.
Pair binding energies for the uniformly distorted
truncated tetrahedron:
(a) Peierls-distorted; (b) Jahn-Teller distorted.
Notice the absence of pairing at small $U$ in (b).
\vglue 3.6truecm
\noindent Fig.~4.
Pair binding energies in the self-consistently
distorted truncated tetrahedron.
Note that the pairing is strongest at $U$=0.
\endinsert

To calculate the effects of variable distortion we use the
Peierls-extended Hubbard Hamiltonian,
$$
H=\sum_{\langle ij\rangle,\sigma}
[-t_0+\alpha\delta_{ij}](c_{i,\sigma}^+c_{j,\sigma}
+c_{j,\sigma}^+c_{i,\sigma})
+U\sum_in_{i,\uparrow}n_{i,\downarrow}
+ {1\over 2}\kappa \sum_{\langle ij\rangle}(\delta_{ij}-d_0)^2
\eqno(2') $$
\noindent
where the $\delta_{ij}$ are now self-consistently
chosen to minimize the total energy, and thus depend on filling.
In Fig.~4 we show the results of our calculation for the
truncated tetrahedron.
$d_0$ and $\kappa$ were chosen to depend on $U$ such that the
minimum energy distortion at $N_{el}$=$N$ was independent of $U$:
$\delta_{ij}$=0.1 on the triangular
faces and $\delta_{ij}$=-0.2 on the bonds connecting triangular faces.
Using these values the self-consistent distortion
for each filling was then found and the pairing as given
in Eq.~1 calculated.  Here, we find a greatly enhanced pairing,
though maximum at $U$=0 indicating again that it is driven by the
Jahn-Teller distortion and not the electron-electron interaction
(``bipolaronic" pairing).
\vskip 12truept \noindent
\noindent {\bf 3. Conclusions}
\vskip 12truept \noindent
We have calculated the pair-binding energy for a number of three
dimensional molecules, and find the following. For an $N$ site
system, pairing at small $U$ occurs only
if the $(N - 1)$  (or $(N + 1)$) electron system
has a strong tendency to have a Jahn-Teller distortion. This
pairing is destroyed when a fixed Jahn-Teller distortion is introduced.
Thus the mechanism of pairing in the undistorted cluster at small $U$
in all these cases is related to the suppression of
Jahn-Teller distortion. Since the Jahn-Teller
distortion results from the discrete level degeneracies in
finite molecules, we conclude that the observed pairing in the
small $U$ region is a finite size effect.

Pairing at large $U$ is unrelated to the tendency to have
Jahn-Teller distortion [4], and is not necessarily accompanied
by pairing at small $U$. The occurrence of such large $U$ pairing
in nonbipartite systems
({\it e.g.}, in a pentagonal prism) indicates that it is
also apparently unrelated to antiferromagnetism.
It may be related to energy level crossings.
Currently we are investigating whether the large $U$ pairing
is also a finite size effect, and
if so, the nature of the finite size effect.

The arguments here and in Ref.~4 indicate the absence of pairing in the two
dimensional Hubbard model, as well as in weakly coupled layers.
However, as C$_{60}$ is a finite molecule, this
``on-ball" pairing mechanism cannot be ruled out.
Sawatzky [5] has argued, though,
that since the long-range Coulomb interaction in a C$_{60}$ molecule
is essentially constant --
for this finite system screening moves charge to
the other side of the ball, not infinity, enhancing
the long range interaction while suppressing the short range
interaction
--  a (multi-orbital) Hubbard model where the ``site"
is a C$_{60}$ in the molecular crystal
rather than a C on an individual ball should be used.
In this case the above conclusion that this pairing mechanism
does not operate in infinite systems holds.


Finally, we note that the definition of pairing in Eq.~1 clearly
is not synonymous with superconducting pairing; it may indicate
{\it e.g.} phase segregation [4,6] instead.
We also have not discussed the consequences of longer range
Coulomb interactions. We note that $V$ has been found to destroy
pairing [2], though potentially it can be restored by
other nearest neighbor Coulomb terms [7].
Clearly the effect of
Jahn-Teller interactions must also be included before such
conclusions can be considered final.

{\it Acknowledgements}.
JTG was supported by a NRC-NRaD Research Associateship
through grants from the ONR and DOE.

\bigskip
\centerline{\bf REFERENCES}
\medskip
\item{1.}
S. Chakravarty, M.P. Gelfand, and S. Kivelson,
{\it Science} {\bf 254}, 970 (1991).
\item{2.}
S.R. White, S. Chakravarty, M.P. Gelfand, and S.A. Kivelson,
{\it Phys. Rev. B} {\bf 45}, 5062 (1992).
\item{3.}
R.M. Fye, M.J. Martins, and R.T. Scalettar,
{\it Phys. Rev. B} {\bf 42}, 6809 (1990).
\item{4.}
F. Guo, D. Guo, K.C. Ung, S. Mazumdar, and J.T. Gammel,
to appear in the Proceedings of the Discussion Meeting on Strongly
Correlated Systems in Chemistry, Bangalore, India, 4-8 January 1993
(Springer) (cond-mat/9303003).
\item{5.}
G.A. Sawatzky, these proceedings.
\item{6.}
V.J. Emery, S.A. Kivelson, and H.Q. Lin,
{\it Phys. Rev. Lett.} {\bf 64}, 475 (1990).
\item{7.}
D.K. Campbell, private communication.

\end

\item{} D. G. Kanhere,
to appear in the Proceedings of the Discussion Meeting on Strongly Correlated
Systems in Chemistry, Bangalore, India, 4-8 January 1993 (Springer).
\item{} J. C. Bonner and M. E. Fisher,
{\it Phys. Rev.} {\bf 135}, A640, (1964).
\item{} B. S. Hudson, B. E. Kohler, and K. Schulten, {\it Excited States}
{\bf 6}, edited by E. C. Lim, (Academic, New York) (1982).
\item{} K. Schulten, I. Ohmine, and M. Karplus,
{\it J. Chem. Phys.} {\bf 64}, 4222, (1976).
\item{} Dandan Guo, S. Mazumdar, S. N. Dixit, {\it Nonlinear Optics}, (1993).
\item{} G. P. Agrawal, C. Cojan, and C. Flytzanis,
{\it Phys. Rev. B} {\bf 17}, 776, (1978).
\item{} B. Srinivasan and S. Ramasesha,
{\it Solid State Commun.} {\bf 81}, 831, (1992).
\item{} Z. G. Soos and S. Ramasesha,
{\it J. Chem. Phys.} {\bf 90}, 1067, (1989).
\item{} W. E. Torruellas, K. B. Rochford, R. Zanoni, S. Aramaki,
and G. I. Stegeman, {\it Opt. Commun.} {\bf 82}, 94, (1991).
\item{} J. B. van Beek, F. Kajzar, and A. C. Albrecht,
{\it J. Chem. Phys.} {\bf 95}, 6400, (1991).
\item{} S. Mazumdar and D. K. Campbell,
{\it Phys. Rev. Lett.} {\bf 55}, 2067, (1985).
\item{} Z. G. Soos and S. Ramasesha,
{\it Phys. Rev. B} {\bf 29}, 5410, (1984).
\item{} R. G. Kepler and Z. G. Soos,
{\it Phys. Rev. B} {\bf 43}, 12530, (1991).
\item{} J. R. G. Thorne, Y. Ohsako, R. M. Hochstrasser, and J. M. Zeigler,
{\it Chem. Phys. Lett.} {\bf 162}, 455, (1989).
\item{} S. Mazumdar and S. Ramasesha,
{\it Synth. Metals} {\bf 27}, A105, (1988).

save 50 dict begin /psplot exch def
/StartPSPlot
   {newpath 0 0 moveto 6 setlinewidth 0 setgray 1 setlinecap
    /imtx matrix currentmatrix def /dmtx matrix defaultmatrix def
    /fnt /Courier findfont def /smtx matrix def fnt 8 scalefont setfont}def
/solid {{}0}def
/dotted	{[2 nail 10 nail ] 0}def
/longdashed {[10 nail] 0}def
/shortdashed {[6 nail] 0}def
/dotdashed {[2 nail 6 nail 10 nail 6 nail] 0}def
/min {2 copy lt{pop}{exch pop}ifelse}def
/max {2 copy lt{exch pop}{pop}ifelse}def
/len {dup mul exch dup mul add sqrt}def
/nail {0 imtx dtransform len 0 idtransform len}def

/m {newpath moveto}def
/n {lineto currentpoint stroke moveto}def
/p {newpath moveto gsave 1 setlinecap solid setdash
    dmtx setmatrix .4 nail setlinewidth
    .05 0 idtransform rlineto stroke grestore}def
/l {moveto lineto currentpoint stroke moveto}def
/t {smtx currentmatrix pop imtx setmatrix show smtx setmatrix}def
/a {gsave newpath /y2 exch def /x2 exch def
    /y1 exch def /x1 exch def /yc exch def /xc exch def
    /r x1 xc sub dup mul y1 yc sub dup mul add sqrt
       x2 xc sub dup mul y2 yc sub dup mul add sqrt add 2 div def
    /ang1 y1 yc sub x1 xc sub atan def
    /ang2 y2 yc sub x2 xc sub atan def
    xc yc r ang1 ang2 arc stroke grestore}def
/c {gsave newpath 0 360 arc stroke grestore}def
/e {gsave showpage grestore newpath 0 0 moveto}def
/f {load exec setdash}def
/s {/ury exch def /urx exch def /lly exch def /llx exch def
    imtx setmatrix newpath clippath pathbbox newpath
    /dury exch def /durx exch def /dlly exch def /dllx exch def
    /md durx dllx sub dury dlly sub min def
    /Mu urx llx sub ury lly sub max def
    dllx dlly translate md Mu div dup scale llx neg lly neg translate}def
/EndPSPlot {clear psplot end restore}def

StartPSPlot
0 0 4096 4096 s
/solid f
/solid f
0 0 4096 4096 s
/solid f
587 4606 m
397 4459 n
533 4311 n
804 4311 n
939 4459 n
749 4606 n
587 4606 n
587 4976 m
397 4828 n
533 4680 n
804 4680 n
939 4828 n
749 4976 n
587 4976 n
587 4606 m
587 4976 n
397 4459 m
397 4828 n
533 4311 m
533 4680 n
804 4311 m
804 4680 n
939 4459 m
939 4828 n
749 4606 m
749 4976 n
/solid f
462 5134 m
456 5128 n
450 5119 n
444 5107 n
441 5092 n
441 5080 n
444 5065 n
450 5053 n
456 5044 n
462 5038 n
456 5128 m
450 5116 n
447 5107 n
444 5092 n
444 5080 n
447 5065 n
450 5056 n
456 5044 n
507 5089 m
486 5095 m
486 5092 n
483 5092 n
483 5095 n
486 5098 n
492 5101 n
504 5101 n
510 5098 n
513 5095 n
516 5089 n
516 5068 n
519 5062 n
522 5059 n
513 5095 m
513 5068 n
516 5062 n
522 5059 n
525 5059 n
513 5089 m
510 5086 n
492 5083 n
483 5080 n
480 5074 n
480 5068 n
483 5062 n
492 5059 n
501 5059 n
507 5062 n
513 5068 n
492 5083 m
486 5080 n
483 5074 n
483 5068 n
486 5062 n
492 5059 n
567 5089 m
540 5134 m
546 5128 n
552 5119 n
558 5107 n
561 5092 n
561 5080 n
558 5065 n
552 5053 n
546 5044 n
540 5038 n
546 5128 m
552 5116 n
555 5107 n
558 5092 n
558 5080 n
555 5065 n
552 5056 n
546 5044 n
609 5089 m
657 5089 m
666 5122 m
666 5059 n
669 5122 m
669 5059 n
666 5092 m
660 5098 n
654 5101 n
648 5101 n
639 5098 n
633 5092 n
630 5083 n
630 5077 n
633 5068 n
639 5062 n
648 5059 n
654 5059 n
660 5062 n
666 5068 n
648 5101 m
642 5098 n
636 5092 n
633 5083 n
633 5077 n
636 5068 n
642 5062 n
648 5059 n
657 5122 m
669 5122 n
666 5059 m
678 5059 n
720 5089 m
699 5101 m
699 5059 n
702 5101 m
702 5059 n
702 5083 m
705 5092 n
711 5098 n
717 5101 n
726 5101 n
729 5098 n
729 5095 n
726 5092 n
723 5095 n
726 5098 n
690 5101 m
702 5101 n
690 5059 m
711 5059 n
771 5089 m
750 5101 m
750 5068 n
753 5062 n
762 5059 n
768 5059 n
777 5062 n
783 5068 n
753 5101 m
753 5068 n
756 5062 n
762 5059 n
783 5101 m
783 5059 n
786 5101 m
786 5059 n
741 5101 m
753 5101 n
774 5101 m
786 5101 n
783 5059 m
795 5059 n
837 5089 m
816 5101 m
816 5059 n
819 5101 m
819 5059 n
819 5092 m
825 5098 n
834 5101 n
840 5101 n
849 5098 n
852 5092 n
852 5059 n
840 5101 m
846 5098 n
849 5092 n
849 5059 n
852 5092 m
858 5098 n
867 5101 n
873 5101 n
882 5098 n
885 5092 n
885 5059 n
873 5101 m
879 5098 n
882 5092 n
882 5059 n
807 5101 m
819 5101 n
807 5059 m
828 5059 n
840 5059 m
861 5059 n
873 5059 m
894 5059 n
0 0 4096 4096 s
/solid f
1562 4606 m
/solid f
/solid f
1562 4606 m
1372 4459 n
1508 4311 n
1779 4311 n
1914 4459 n
1724 4606 n
/shortdashed f
1724 4606 m
1562 4976 n
/solid f
1562 4976 m
1372 4828 n
1508 4680 n
1779 4680 n
1914 4828 n
1724 4976 n
1562 4606 n
1562 4606 m
1562 4976 n
1372 4459 m
1372 4828 n
1508 4311 m
1508 4680 n
1779 4311 m
1779 4680 n
1914 4459 m
1914 4828 n
1724 4606 m
1724 4976 n
/solid f
1229 5134 m
1223 5128 n
1217 5119 n
1211 5107 n
1208 5092 n
1208 5080 n
1211 5065 n
1217 5053 n
1223 5044 n
1229 5038 n
1223 5128 m
1217 5116 n
1214 5107 n
1211 5092 n
1211 5080 n
1214 5065 n
1217 5056 n
1223 5044 n
1274 5089 m
1253 5122 m
1253 5059 n
1256 5122 m
1256 5059 n
1256 5092 m
1262 5098 n
1268 5101 n
1274 5101 n
1283 5098 n
1289 5092 n
1292 5083 n
1292 5077 n
1289 5068 n
1283 5062 n
1274 5059 n
1268 5059 n
1262 5062 n
1256 5068 n
1274 5101 m
1280 5098 n
1286 5092 n
1289 5083 n
1289 5077 n
1286 5068 n
1280 5062 n
1274 5059 n
1244 5122 m
1256 5122 n
1337 5089 m
1310 5134 m
1316 5128 n
1322 5119 n
1328 5107 n
1331 5092 n
1331 5080 n
1328 5065 n
1322 5053 n
1316 5044 n
1310 5038 n
1316 5128 m
1322 5116 n
1325 5107 n
1328 5092 n
1328 5080 n
1325 5065 n
1322 5056 n
1316 5044 n
1379 5089 m
1427 5089 m
1406 5122 m
1406 5071 n
1409 5062 n
1415 5059 n
1421 5059 n
1427 5062 n
1430 5068 n
1409 5122 m
1409 5071 n
1412 5062 n
1415 5059 n
1397 5101 m
1421 5101 n
1472 5089 m
1448 5101 m
1460 5059 n
1451 5101 m
1460 5068 n
1472 5101 m
1460 5059 n
1472 5101 m
1484 5059 n
1475 5101 m
1484 5068 n
1496 5101 m
1484 5059 n
1439 5101 m
1460 5101 n
1487 5101 m
1505 5101 n
1544 5089 m
1523 5122 m
1520 5119 n
1523 5116 n
1526 5119 n
1523 5122 n
1523 5101 m
1523 5059 n
1526 5101 m
1526 5059 n
1514 5101 m
1526 5101 n
1514 5059 m
1535 5059 n
1577 5089 m
1580 5095 m
1583 5101 n
1583 5089 n
1580 5095 n
1577 5098 n
1571 5101 n
1559 5101 n
1553 5098 n
1550 5095 n
1550 5089 n
1553 5086 n
1559 5083 n
1574 5077 n
1580 5074 n
1583 5071 n
1550 5092 m
1553 5089 n
1559 5086 n
1574 5080 n
1580 5077 n
1583 5074 n
1583 5065 n
1580 5062 n
1574 5059 n
1562 5059 n
1556 5062 n
1553 5065 n
1550 5071 n
1550 5059 n
1553 5065 n
1628 5089 m
1607 5122 m
1607 5071 n
1610 5062 n
1616 5059 n
1622 5059 n
1628 5062 n
1631 5068 n
1610 5122 m
1610 5071 n
1613 5062 n
1616 5059 n
1598 5101 m
1622 5101 n
1673 5089 m
1649 5083 m
1685 5083 n
1685 5089 n
1682 5095 n
1679 5098 n
1673 5101 n
1664 5101 n
1655 5098 n
1649 5092 n
1646 5083 n
1646 5077 n
1649 5068 n
1655 5062 n
1664 5059 n
1670 5059 n
1679 5062 n
1685 5068 n
1682 5083 m
1682 5092 n
1679 5098 n
1664 5101 m
1658 5098 n
1652 5092 n
1649 5083 n
1649 5077 n
1652 5068 n
1658 5062 n
1664 5059 n
1730 5089 m
1739 5122 m
1739 5059 n
1742 5122 m
1742 5059 n
1739 5092 m
1733 5098 n
1727 5101 n
1721 5101 n
1712 5098 n
1706 5092 n
1703 5083 n
1703 5077 n
1706 5068 n
1712 5062 n
1721 5059 n
1727 5059 n
1733 5062 n
1739 5068 n
1721 5101 m
1715 5098 n
1709 5092 n
1706 5083 n
1706 5077 n
1709 5068 n
1715 5062 n
1721 5059 n
1730 5122 m
1742 5122 n
1739 5059 m
1751 5059 n
1793 5089 m
1841 5089 m
1850 5122 m
1850 5059 n
1853 5122 m
1853 5059 n
1850 5092 m
1844 5098 n
1838 5101 n
1832 5101 n
1823 5098 n
1817 5092 n
1814 5083 n
1814 5077 n
1817 5068 n
1823 5062 n
1832 5059 n
1838 5059 n
1844 5062 n
1850 5068 n
1832 5101 m
1826 5098 n
1820 5092 n
1817 5083 n
1817 5077 n
1820 5068 n
1826 5062 n
1832 5059 n
1841 5122 m
1853 5122 n
1850 5059 m
1862 5059 n
1904 5089 m
1883 5101 m
1883 5059 n
1886 5101 m
1886 5059 n
1886 5083 m
1889 5092 n
1895 5098 n
1901 5101 n
1910 5101 n
1913 5098 n
1913 5095 n
1910 5092 n
1907 5095 n
1910 5098 n
1874 5101 m
1886 5101 n
1874 5059 m
1895 5059 n
1955 5089 m
1934 5101 m
1934 5068 n
1937 5062 n
1946 5059 n
1952 5059 n
1961 5062 n
1967 5068 n
1937 5101 m
1937 5068 n
1940 5062 n
1946 5059 n
1967 5101 m
1967 5059 n
1970 5101 m
1970 5059 n
1925 5101 m
1937 5101 n
1958 5101 m
1970 5101 n
1967 5059 m
1979 5059 n
2021 5089 m
2000 5101 m
2000 5059 n
2003 5101 m
2003 5059 n
2003 5092 m
2009 5098 n
2018 5101 n
2024 5101 n
2033 5098 n
2036 5092 n
2036 5059 n
2024 5101 m
2030 5098 n
2033 5092 n
2033 5059 n
2036 5092 m
2042 5098 n
2051 5101 n
2057 5101 n
2066 5098 n
2069 5092 n
2069 5059 n
2057 5101 m
2063 5098 n
2066 5092 n
2066 5059 n
1991 5101 m
2003 5101 n
1991 5059 m
2012 5059 n
2024 5059 m
2045 5059 n
2057 5059 m
2078 5059 n
0 0 4096 4096 s
/solid f
2862 4508 m
/solid f
/solid f
2808 4291 m
3079 4291 n
2943 4483 n
2808 4291 n
2943 4754 m
2808 4947 n
3079 4947 n
2943 4754 n
2808 4619 m
2616 4754 n
2616 4483 n
2808 4619 n
3079 4619 m
3271 4754 n
3271 4483 n
3079 4619 n
/shortdashed f
2943 4754 m
2943 4483 n
/solid f
2808 4291 m
2616 4483 n
3271 4483 m
3079 4291 n
2808 4947 m
2616 4754 n
3079 4947 m
3271 4754 n
2808 4619 m
3079 4619 n
/solid f
2289 5134 m
2283 5128 n
2277 5119 n
2271 5107 n
2268 5092 n
2268 5080 n
2271 5065 n
2277 5053 n
2283 5044 n
2289 5038 n
2283 5128 m
2277 5116 n
2274 5107 n
2271 5092 n
2271 5080 n
2274 5065 n
2277 5056 n
2283 5044 n
2334 5089 m
2343 5092 m
2340 5089 n
2343 5086 n
2346 5089 n
2346 5092 n
2340 5098 n
2334 5101 n
2325 5101 n
2316 5098 n
2310 5092 n
2307 5083 n
2307 5077 n
2310 5068 n
2316 5062 n
2325 5059 n
2331 5059 n
2340 5062 n
2346 5068 n
2325 5101 m
2319 5098 n
2313 5092 n
2310 5083 n
2310 5077 n
2313 5068 n
2319 5062 n
2325 5059 n
2391 5089 m
2364 5134 m
2370 5128 n
2376 5119 n
2382 5107 n
2385 5092 n
2385 5080 n
2382 5065 n
2376 5053 n
2370 5044 n
2364 5038 n
2370 5128 m
2376 5116 n
2379 5107 n
2382 5092 n
2382 5080 n
2379 5065 n
2376 5056 n
2370 5044 n
2433 5089 m
2481 5089 m
2460 5122 m
2460 5071 n
2463 5062 n
2469 5059 n
2475 5059 n
2481 5062 n
2484 5068 n
2463 5122 m
2463 5071 n
2466 5062 n
2469 5059 n
2451 5101 m
2475 5101 n
2526 5089 m
2505 5101 m
2505 5059 n
2508 5101 m
2508 5059 n
2508 5083 m
2511 5092 n
2517 5098 n
2523 5101 n
2532 5101 n
2535 5098 n
2535 5095 n
2532 5092 n
2529 5095 n
2532 5098 n
2496 5101 m
2508 5101 n
2496 5059 m
2517 5059 n
2577 5089 m
2556 5101 m
2556 5068 n
2559 5062 n
2568 5059 n
2574 5059 n
2583 5062 n
2589 5068 n
2559 5101 m
2559 5068 n
2562 5062 n
2568 5059 n
2589 5101 m
2589 5059 n
2592 5101 m
2592 5059 n
2547 5101 m
2559 5101 n
2580 5101 m
2592 5101 n
2589 5059 m
2601 5059 n
2643 5089 m
2622 5101 m
2622 5059 n
2625 5101 m
2625 5059 n
2625 5092 m
2631 5098 n
2640 5101 n
2646 5101 n
2655 5098 n
2658 5092 n
2658 5059 n
2646 5101 m
2652 5098 n
2655 5092 n
2655 5059 n
2613 5101 m
2625 5101 n
2613 5059 m
2634 5059 n
2646 5059 m
2667 5059 n
2709 5089 m
2718 5092 m
2715 5089 n
2718 5086 n
2721 5089 n
2721 5092 n
2715 5098 n
2709 5101 n
2700 5101 n
2691 5098 n
2685 5092 n
2682 5083 n
2682 5077 n
2685 5068 n
2691 5062 n
2700 5059 n
2706 5059 n
2715 5062 n
2721 5068 n
2700 5101 m
2694 5098 n
2688 5092 n
2685 5083 n
2685 5077 n
2688 5068 n
2694 5062 n
2700 5059 n
2766 5089 m
2745 5095 m
2745 5092 n
2742 5092 n
2742 5095 n
2745 5098 n
2751 5101 n
2763 5101 n
2769 5098 n
2772 5095 n
2775 5089 n
2775 5068 n
2778 5062 n
2781 5059 n
2772 5095 m
2772 5068 n
2775 5062 n
2781 5059 n
2784 5059 n
2772 5089 m
2769 5086 n
2751 5083 n
2742 5080 n
2739 5074 n
2739 5068 n
2742 5062 n
2751 5059 n
2760 5059 n
2766 5062 n
2772 5068 n
2751 5083 m
2745 5080 n
2742 5074 n
2742 5068 n
2745 5062 n
2751 5059 n
2826 5089 m
2805 5122 m
2805 5071 n
2808 5062 n
2814 5059 n
2820 5059 n
2826 5062 n
2829 5068 n
2808 5122 m
2808 5071 n
2811 5062 n
2814 5059 n
2796 5101 m
2820 5101 n
2871 5089 m
2847 5083 m
2883 5083 n
2883 5089 n
2880 5095 n
2877 5098 n
2871 5101 n
2862 5101 n
2853 5098 n
2847 5092 n
2844 5083 n
2844 5077 n
2847 5068 n
2853 5062 n
2862 5059 n
2868 5059 n
2877 5062 n
2883 5068 n
2880 5083 m
2880 5092 n
2877 5098 n
2862 5101 m
2856 5098 n
2850 5092 n
2847 5083 n
2847 5077 n
2850 5068 n
2856 5062 n
2862 5059 n
2928 5089 m
2937 5122 m
2937 5059 n
2940 5122 m
2940 5059 n
2937 5092 m
2931 5098 n
2925 5101 n
2919 5101 n
2910 5098 n
2904 5092 n
2901 5083 n
2901 5077 n
2904 5068 n
2910 5062 n
2919 5059 n
2925 5059 n
2931 5062 n
2937 5068 n
2919 5101 m
2913 5098 n
2907 5092 n
2904 5083 n
2904 5077 n
2907 5068 n
2913 5062 n
2919 5059 n
2928 5122 m
2940 5122 n
2937 5059 m
2949 5059 n
2991 5089 m
3039 5089 m
3018 5122 m
3018 5071 n
3021 5062 n
3027 5059 n
3033 5059 n
3039 5062 n
3042 5068 n
3021 5122 m
3021 5071 n
3024 5062 n
3027 5059 n
3009 5101 m
3033 5101 n
3084 5089 m
3060 5083 m
3096 5083 n
3096 5089 n
3093 5095 n
3090 5098 n
3084 5101 n
3075 5101 n
3066 5098 n
3060 5092 n
3057 5083 n
3057 5077 n
3060 5068 n
3066 5062 n
3075 5059 n
3081 5059 n
3090 5062 n
3096 5068 n
3093 5083 m
3093 5092 n
3090 5098 n
3075 5101 m
3069 5098 n
3063 5092 n
3060 5083 n
3060 5077 n
3063 5068 n
3069 5062 n
3075 5059 n
3141 5089 m
3120 5122 m
3120 5071 n
3123 5062 n
3129 5059 n
3135 5059 n
3141 5062 n
3144 5068 n
3123 5122 m
3123 5071 n
3126 5062 n
3129 5059 n
3111 5101 m
3135 5101 n
3186 5089 m
3165 5101 m
3165 5059 n
3168 5101 m
3168 5059 n
3168 5083 m
3171 5092 n
3177 5098 n
3183 5101 n
3192 5101 n
3195 5098 n
3195 5095 n
3192 5092 n
3189 5095 n
3192 5098 n
3156 5101 m
3168 5101 n
3156 5059 m
3177 5059 n
3237 5089 m
3216 5095 m
3216 5092 n
3213 5092 n
3213 5095 n
3216 5098 n
3222 5101 n
3234 5101 n
3240 5098 n
3243 5095 n
3246 5089 n
3246 5068 n
3249 5062 n
3252 5059 n
3243 5095 m
3243 5068 n
3246 5062 n
3252 5059 n
3255 5059 n
3243 5089 m
3240 5086 n
3222 5083 n
3213 5080 n
3210 5074 n
3210 5068 n
3213 5062 n
3222 5059 n
3231 5059 n
3237 5062 n
3243 5068 n
3222 5083 m
3216 5080 n
3213 5074 n
3213 5068 n
3216 5062 n
3222 5059 n
3297 5089 m
3276 5122 m
3276 5059 n
3279 5122 m
3279 5059 n
3279 5092 m
3285 5098 n
3294 5101 n
3300 5101 n
3309 5098 n
3312 5092 n
3312 5059 n
3300 5101 m
3306 5098 n
3309 5092 n
3309 5059 n
3267 5122 m
3279 5122 n
3267 5059 m
3288 5059 n
3300 5059 m
3321 5059 n
3363 5089 m
3339 5083 m
3375 5083 n
3375 5089 n
3372 5095 n
3369 5098 n
3363 5101 n
3354 5101 n
3345 5098 n
3339 5092 n
3336 5083 n
3336 5077 n
3339 5068 n
3345 5062 n
3354 5059 n
3360 5059 n
3369 5062 n
3375 5068 n
3372 5083 m
3372 5092 n
3369 5098 n
3354 5101 m
3348 5098 n
3342 5092 n
3339 5083 n
3339 5077 n
3342 5068 n
3348 5062 n
3354 5059 n
3420 5089 m
3429 5122 m
3429 5059 n
3432 5122 m
3432 5059 n
3429 5092 m
3423 5098 n
3417 5101 n
3411 5101 n
3402 5098 n
3396 5092 n
3393 5083 n
3393 5077 n
3396 5068 n
3402 5062 n
3411 5059 n
3417 5059 n
3423 5062 n
3429 5068 n
3411 5101 m
3405 5098 n
3399 5092 n
3396 5083 n
3396 5077 n
3399 5068 n
3405 5062 n
3411 5059 n
3420 5122 m
3432 5122 n
3429 5059 m
3441 5059 n
3483 5089 m
3462 5101 m
3462 5059 n
3465 5101 m
3465 5059 n
3465 5083 m
3468 5092 n
3474 5098 n
3480 5101 n
3489 5101 n
3492 5098 n
3492 5095 n
3489 5092 n
3486 5095 n
3489 5098 n
3453 5101 m
3465 5101 n
3453 5059 m
3474 5059 n
3534 5089 m
3525 5101 m
3516 5098 n
3510 5092 n
3507 5083 n
3507 5077 n
3510 5068 n
3516 5062 n
3525 5059 n
3531 5059 n
3540 5062 n
3546 5068 n
3549 5077 n
3549 5083 n
3546 5092 n
3540 5098 n
3531 5101 n
3525 5101 n
3525 5101 m
3519 5098 n
3513 5092 n
3510 5083 n
3510 5077 n
3513 5068 n
3519 5062 n
3525 5059 n
3531 5059 m
3537 5062 n
3543 5068 n
3546 5077 n
3546 5083 n
3543 5092 n
3537 5098 n
3531 5101 n
3594 5089 m
3573 5101 m
3573 5059 n
3576 5101 m
3576 5059 n
3576 5092 m
3582 5098 n
3591 5101 n
3597 5101 n
3606 5098 n
3609 5092 n
3609 5059 n
3597 5101 m
3603 5098 n
3606 5092 n
3606 5059 n
3564 5101 m
3576 5101 n
3564 5059 m
3585 5059 n
3597 5059 m
3618 5059 n
0 0 4096 4096 s
/solid f
/solid f
883 3608 p
883 3608 p
883 3608 p
883 3608 p
883 3608 p
883 3608 p
883 3608 p
883 3608 p
883 3608 p
883 3608 p
883 3608 p
883 3608 p
883 3608 p
883 3608 p
883 3608 p
883 3608 p
883 3608 p
883 3608 p
883 3608 p
883 3608 p
883 3608 p
883 3608 p
883 3608 p
883 3608 p
883 3608 p
883 3608 p
883 3608 p
883 3608 p
883 3608 p
883 3608 p
883 3608 p
883 3608 p
883 3608 p
883 3608 p
883 3608 p
883 3608 p
883 3608 p
883 3608 p
883 3608 p
883 3608 p
883 3608 p
883 3608 p
883 3608 p
883 3608 p
883 3608 p
883 3608 p
883 3608 p
883 3608 p
883 3608 p
883 3608 p
883 3608 p
883 3608 p
883 3608 p
883 3608 p
883 3608 p
883 3608 p
883 3608 p
883 3608 p
883 3608 p
883 3608 p
883 3608 p
883 3608 p
883 3608 p
883 3608 p
883 3608 p
883 3608 p
883 3608 p
883 3608 p
883 3608 p
883 3608 p
883 3608 p
883 3608 p
883 3608 p
883 3608 p
883 3608 p
883 3608 p
883 3608 p
883 3608 p
883 3608 p
883 3608 p
/shortdashed f
883 3608 m
1016 3549 n
1149 3500 n
1282 3562 n
1415 3685 n
1548 3747 n
1681 3708 n
1814 3618 n
1947 3529 n
2080 3468 n
2212 3439 n
2345 3437 n
2478 3456 n
2611 3492 n
2744 3537 n
2877 3590 n
3010 3646 n
3143 3705 n
3276 3765 n
3409 3824 n
3542 3883 n
/solid f
883 3608 m
1016 3549 n
1149 3500 n
1282 3562 n
1415 3685 n
1548 3747 n
1681 3709 n
1814 3618 n
1947 3529 n
2080 3468 n
2212 3439 n
2345 3437 n
2478 3456 n
2611 3492 n
2744 3537 n
2877 3590 n
3010 3646 n
3143 3705 n
3276 3765 n
3409 3824 n
3542 3883 n
/solid f
750 3425 m
3675 3425 n
3675 3912 n
750 3912 n
750 3425 n
883 3425 m
883 3525 n
883 3812 m
883 3912 n
1149 3425 m
1149 3475 n
1149 3862 m
1149 3912 n
1415 3425 m
1415 3475 n
1415 3862 m
1415 3912 n
1681 3425 m
1681 3475 n
1681 3862 m
1681 3912 n
1947 3425 m
1947 3475 n
1947 3862 m
1947 3912 n
2212 3425 m
2212 3525 n
2212 3812 m
2212 3912 n
2478 3425 m
2478 3475 n
2478 3862 m
2478 3912 n
2744 3425 m
2744 3475 n
2744 3862 m
2744 3912 n
3010 3425 m
3010 3475 n
3010 3862 m
3010 3912 n
3276 3425 m
3276 3475 n
3276 3862 m
3276 3912 n
3542 3425 m
3542 3525 n
3542 3812 m
3542 3912 n
750 3437 m
800 3437 n
3625 3437 m
3675 3437 n
750 3462 m
800 3462 n
3625 3462 m
3675 3462 n
750 3486 m
850 3486 n
3575 3486 m
3675 3486 n
750 3510 m
800 3510 n
3625 3510 m
3675 3510 n
750 3535 m
800 3535 n
3625 3535 m
3675 3535 n
750 3559 m
800 3559 n
3625 3559 m
3675 3559 n
750 3583 m
800 3583 n
3625 3583 m
3675 3583 n
750 3608 m
850 3608 n
3575 3608 m
3675 3608 n
750 3632 m
800 3632 n
3625 3632 m
3675 3632 n
750 3657 m
800 3657 n
3625 3657 m
3675 3657 n
750 3681 m
800 3681 n
3625 3681 m
3675 3681 n
750 3705 m
800 3705 n
3625 3705 m
3675 3705 n
750 3730 m
850 3730 n
3575 3730 m
3675 3730 n
750 3754 m
800 3754 n
3625 3754 m
3675 3754 n
750 3778 m
800 3778 n
3625 3778 m
3675 3778 n
750 3803 m
800 3803 n
3625 3803 m
3675 3803 n
750 3827 m
800 3827 n
3625 3827 m
3675 3827 n
750 3852 m
850 3852 n
3575 3852 m
3675 3852 n
750 3876 m
800 3876 n
3625 3876 m
3675 3876 n
750 3900 m
800 3900 n
3625 3900 m
3675 3900 n
942 3883 m
936 3877 n
930 3868 n
924 3856 n
921 3841 n
921 3829 n
924 3814 n
930 3802 n
936 3793 n
942 3787 n
936 3877 m
930 3865 n
927 3856 n
924 3841 n
924 3829 n
927 3814 n
930 3805 n
936 3793 n
987 3838 m
966 3844 m
966 3841 n
963 3841 n
963 3844 n
966 3847 n
972 3850 n
984 3850 n
990 3847 n
993 3844 n
996 3838 n
996 3817 n
999 3811 n
1002 3808 n
993 3844 m
993 3817 n
996 3811 n
1002 3808 n
1005 3808 n
993 3838 m
990 3835 n
972 3832 n
963 3829 n
960 3823 n
960 3817 n
963 3811 n
972 3808 n
981 3808 n
987 3811 n
993 3817 n
972 3832 m
966 3829 n
963 3823 n
963 3817 n
966 3811 n
972 3808 n
1047 3838 m
1020 3883 m
1026 3877 n
1032 3868 n
1038 3856 n
1041 3841 n
1041 3829 n
1038 3814 n
1032 3802 n
1026 3793 n
1020 3787 n
1026 3877 m
1032 3865 n
1035 3856 n
1038 3841 n
1038 3829 n
1035 3814 n
1032 3805 n
1026 3793 n
433 3483 m
487 3483 n
535 3486 m
526 3519 m
517 3516 n
511 3507 n
508 3492 n
508 3483 n
511 3468 n
517 3459 n
526 3456 n
532 3456 n
541 3459 n
547 3468 n
550 3483 n
550 3492 n
547 3507 n
541 3516 n
532 3519 n
526 3519 n
526 3519 m
520 3516 n
517 3513 n
514 3507 n
511 3492 n
511 3483 n
514 3468 n
517 3462 n
520 3459 n
526 3456 n
532 3456 m
538 3459 n
541 3462 n
544 3468 n
547 3483 n
547 3492 n
544 3507 n
541 3513 n
538 3516 n
532 3519 n
595 3486 m
574 3462 m
571 3459 n
574 3456 n
577 3459 n
574 3462 n
625 3486 m
616 3519 m
607 3516 n
601 3507 n
598 3492 n
598 3483 n
601 3468 n
607 3459 n
616 3456 n
622 3456 n
631 3459 n
637 3468 n
640 3483 n
640 3492 n
637 3507 n
631 3516 n
622 3519 n
616 3519 n
616 3519 m
610 3516 n
607 3513 n
604 3507 n
601 3492 n
601 3483 n
604 3468 n
607 3462 n
610 3459 n
616 3456 n
622 3456 m
628 3459 n
631 3462 n
634 3468 n
637 3483 n
637 3492 n
634 3507 n
631 3513 n
628 3516 n
622 3519 n
685 3486 m
661 3507 m
664 3504 n
661 3501 n
658 3504 n
658 3507 n
661 3513 n
664 3516 n
673 3519 n
685 3519 n
694 3516 n
697 3513 n
700 3507 n
700 3501 n
697 3495 n
688 3489 n
673 3483 n
667 3480 n
661 3474 n
658 3465 n
658 3456 n
685 3519 m
691 3516 n
694 3513 n
697 3507 n
697 3501 n
694 3495 n
685 3489 n
673 3483 n
658 3462 m
661 3465 n
667 3465 n
682 3459 n
691 3459 n
697 3462 n
700 3465 n
667 3465 m
682 3456 n
694 3456 n
697 3459 n
700 3465 n
700 3471 n
676 3641 m
667 3638 n
661 3629 n
658 3614 n
658 3605 n
661 3590 n
667 3581 n
676 3578 n
682 3578 n
691 3581 n
697 3590 n
700 3605 n
700 3614 n
697 3629 n
691 3638 n
682 3641 n
676 3641 n
676 3641 m
670 3638 n
667 3635 n
664 3629 n
661 3614 n
661 3605 n
664 3590 n
667 3584 n
670 3581 n
676 3578 n
682 3578 m
688 3581 n
691 3584 n
694 3590 n
697 3605 n
697 3614 n
694 3629 n
691 3635 n
688 3638 n
682 3641 n
526 3763 m
517 3760 n
511 3751 n
508 3736 n
508 3727 n
511 3712 n
517 3703 n
526 3700 n
532 3700 n
541 3703 n
547 3712 n
550 3727 n
550 3736 n
547 3751 n
541 3760 n
532 3763 n
526 3763 n
526 3763 m
520 3760 n
517 3757 n
514 3751 n
511 3736 n
511 3727 n
514 3712 n
517 3706 n
520 3703 n
526 3700 n
532 3700 m
538 3703 n
541 3706 n
544 3712 n
547 3727 n
547 3736 n
544 3751 n
541 3757 n
538 3760 n
532 3763 n
595 3730 m
574 3706 m
571 3703 n
574 3700 n
577 3703 n
574 3706 n
625 3730 m
616 3763 m
607 3760 n
601 3751 n
598 3736 n
598 3727 n
601 3712 n
607 3703 n
616 3700 n
622 3700 n
631 3703 n
637 3712 n
640 3727 n
640 3736 n
637 3751 n
631 3760 n
622 3763 n
616 3763 n
616 3763 m
610 3760 n
607 3757 n
604 3751 n
601 3736 n
601 3727 n
604 3712 n
607 3706 n
610 3703 n
616 3700 n
622 3700 m
628 3703 n
631 3706 n
634 3712 n
637 3727 n
637 3736 n
634 3751 n
631 3757 n
628 3760 n
622 3763 n
685 3730 m
661 3751 m
664 3748 n
661 3745 n
658 3748 n
658 3751 n
661 3757 n
664 3760 n
673 3763 n
685 3763 n
694 3760 n
697 3757 n
700 3751 n
700 3745 n
697 3739 n
688 3733 n
673 3727 n
667 3724 n
661 3718 n
658 3709 n
658 3700 n
685 3763 m
691 3760 n
694 3757 n
697 3751 n
697 3745 n
694 3739 n
685 3733 n
673 3727 n
658 3706 m
661 3709 n
667 3709 n
682 3703 n
691 3703 n
697 3706 n
700 3709 n
667 3709 m
682 3700 n
694 3700 n
697 3703 n
700 3709 n
700 3715 n
523 3885 m
514 3882 n
508 3873 n
505 3858 n
505 3849 n
508 3834 n
514 3825 n
523 3822 n
529 3822 n
538 3825 n
544 3834 n
547 3849 n
547 3858 n
544 3873 n
538 3882 n
529 3885 n
523 3885 n
523 3885 m
517 3882 n
514 3879 n
511 3873 n
508 3858 n
508 3849 n
511 3834 n
514 3828 n
517 3825 n
523 3822 n
529 3822 m
535 3825 n
538 3828 n
541 3834 n
544 3849 n
544 3858 n
541 3873 n
538 3879 n
535 3882 n
529 3885 n
592 3852 m
571 3828 m
568 3825 n
571 3822 n
574 3825 n
571 3828 n
622 3852 m
613 3885 m
604 3882 n
598 3873 n
595 3858 n
595 3849 n
598 3834 n
604 3825 n
613 3822 n
619 3822 n
628 3825 n
634 3834 n
637 3849 n
637 3858 n
634 3873 n
628 3882 n
619 3885 n
613 3885 n
613 3885 m
607 3882 n
604 3879 n
601 3873 n
598 3858 n
598 3849 n
601 3834 n
604 3828 n
607 3825 n
613 3822 n
619 3822 m
625 3825 n
628 3828 n
631 3834 n
634 3849 n
634 3858 n
631 3873 n
628 3879 n
625 3882 n
619 3885 n
682 3852 m
682 3879 m
682 3822 n
685 3885 m
685 3822 n
685 3885 m
652 3840 n
700 3840 n
673 3822 m
694 3822 n
287 3638 m
350 3614 n
287 3638 m
350 3662 n
296 3638 m
350 3659 n
347 3617 m
347 3659 n
350 3614 m
350 3662 n
320 3704 m
287 3683 m
350 3683 n
287 3686 m
350 3686 n
305 3704 m
329 3704 n
287 3674 m
287 3722 n
305 3722 n
287 3719 n
317 3686 m
317 3704 n
350 3674 m
350 3722 n
332 3722 n
350 3719 n
0 0 4096 4096 s
/solid f
/solid f
883 2937 p
883 2937 p
883 2937 p
883 2937 p
883 2937 p
883 2937 p
883 2937 p
883 2937 p
883 2937 p
883 2937 p
883 2937 p
883 2937 p
883 2937 p
883 2937 p
883 2937 p
883 2937 p
883 2937 p
883 2937 p
883 2937 p
883 2937 p
883 2937 p
883 2937 p
883 2937 p
883 2937 p
883 2937 p
883 2937 p
883 2937 p
883 2937 p
883 2937 p
883 2937 p
883 2937 p
883 2937 p
883 2937 p
883 2937 p
883 2937 p
883 2937 p
883 2937 p
883 2937 p
883 2937 p
883 2937 p
883 2937 p
883 2937 p
883 2937 p
883 2937 p
883 2937 p
883 2937 p
883 2937 p
883 2937 p
883 2937 p
883 2937 p
883 2937 p
883 2937 p
883 2937 p
883 2937 p
883 2937 p
883 2937 p
883 2937 p
883 2937 p
883 2937 p
883 2937 p
883 2937 p
883 2937 p
883 2937 p
883 2937 p
883 2937 p
883 2937 p
883 2937 p
883 2937 p
883 2937 p
883 2937 p
883 2937 p
883 2937 p
883 2937 p
883 2937 p
883 2937 p
883 2937 p
883 2937 p
883 2937 p
883 2937 p
883 2937 p
883 2937 p
883 2937 p
883 2937 p
883 2937 p
883 2937 p
883 2937 p
883 2937 p
883 2937 p
883 2937 p
883 2937 p
883 2937 p
883 2937 p
883 2937 p
883 2937 p
883 2937 p
883 2937 p
883 2937 p
883 2937 p
883 2937 p
883 2937 p
883 2937 p
883 2937 p
883 2937 p
883 2937 p
883 2937 p
883 2937 p
883 2937 p
883 2937 p
883 2937 p
883 2937 p
883 2937 p
883 2937 p
883 2937 p
883 2937 p
883 2937 p
883 2937 p
883 2937 p
883 2937 p
883 2937 p
883 2937 p
883 2937 p
883 2937 p
/shortdashed f
883 2937 m
1016 3154 n
1149 3371 n
1282 3320 n
1415 3198 n
1548 3092 n
1681 3022 n
1814 2985 n
1947 2974 n
2080 2981 n
2212 3000 n
2345 3028 n
2478 3060 n
2611 3095 n
2744 3133 n
2877 3171 n
3010 3209 n
3143 3247 n
3276 3285 n
3409 3323 n
3542 3360 n
/solid f
883 2937 m
1016 3168 n
1149 3406 n
1282 3364 n
1415 3239 n
1548 3124 n
1681 3042 n
1814 2994 n
1947 2973 n
2080 2970 n
2212 2979 n
2345 2996 n
2478 3017 n
2611 3040 n
2744 3065 n
2877 3089 n
3010 3114 n
3143 3137 n
3276 3160 n
3409 3182 n
3542 3203 n
/solid f
750 2937 m
3675 2937 n
3675 3424 n
750 3424 n
750 2937 n
883 2937 m
883 3037 n
883 3324 m
883 3424 n
1149 2937 m
1149 2987 n
1149 3374 m
1149 3424 n
1415 2937 m
1415 2987 n
1415 3374 m
1415 3424 n
1681 2937 m
1681 2987 n
1681 3374 m
1681 3424 n
1947 2937 m
1947 2987 n
1947 3374 m
1947 3424 n
2212 2937 m
2212 3037 n
2212 3324 m
2212 3424 n
2478 2937 m
2478 2987 n
2478 3374 m
2478 3424 n
2744 2937 m
2744 2987 n
2744 3374 m
2744 3424 n
3010 2937 m
3010 2987 n
3010 3374 m
3010 3424 n
3276 2937 m
3276 2987 n
3276 3374 m
3276 3424 n
3542 2937 m
3542 3037 n
3542 3324 m
3542 3424 n
750 2976 m
800 2976 n
3625 2976 m
3675 2976 n
750 3015 m
800 3015 n
3625 3015 m
3675 3015 n
750 3054 m
800 3054 n
3625 3054 m
3675 3054 n
750 3093 m
800 3093 n
3625 3093 m
3675 3093 n
750 3132 m
850 3132 n
3575 3132 m
3675 3132 n
750 3171 m
800 3171 n
3625 3171 m
3675 3171 n
750 3210 m
800 3210 n
3625 3210 m
3675 3210 n
750 3249 m
800 3249 n
3625 3249 m
3675 3249 n
750 3288 m
800 3288 n
3625 3288 m
3675 3288 n
750 3327 m
850 3327 n
3575 3327 m
3675 3327 n
750 3366 m
800 3366 n
3625 3366 m
3675 3366 n
750 3405 m
800 3405 n
3625 3405 m
3675 3405 n
940 3395 m
934 3389 n
928 3380 n
922 3368 n
919 3353 n
919 3341 n
922 3326 n
928 3314 n
934 3305 n
940 3299 n
934 3389 m
928 3377 n
925 3368 n
922 3353 n
922 3341 n
925 3326 n
928 3317 n
934 3305 n
985 3350 m
964 3383 m
964 3320 n
967 3383 m
967 3320 n
967 3353 m
973 3359 n
979 3362 n
985 3362 n
994 3359 n
1000 3353 n
1003 3344 n
1003 3338 n
1000 3329 n
994 3323 n
985 3320 n
979 3320 n
973 3323 n
967 3329 n
985 3362 m
991 3359 n
997 3353 n
1000 3344 n
1000 3338 n
997 3329 n
991 3323 n
985 3320 n
955 3383 m
967 3383 n
1048 3350 m
1021 3395 m
1027 3389 n
1033 3380 n
1039 3368 n
1042 3353 n
1042 3341 n
1039 3326 n
1033 3314 n
1027 3305 n
1021 3299 n
1027 3389 m
1033 3377 n
1036 3368 n
1039 3353 n
1039 3341 n
1036 3326 n
1033 3317 n
1027 3305 n
676 2970 m
667 2967 n
661 2958 n
658 2943 n
658 2934 n
661 2919 n
667 2910 n
676 2907 n
682 2907 n
691 2910 n
697 2919 n
700 2934 n
700 2943 n
697 2958 n
691 2967 n
682 2970 n
676 2970 n
676 2970 m
670 2967 n
667 2964 n
664 2958 n
661 2943 n
661 2934 n
664 2919 n
667 2913 n
670 2910 n
676 2907 n
682 2907 m
688 2910 n
691 2913 n
694 2919 n
697 2934 n
697 2943 n
694 2958 n
691 2964 n
688 2967 n
682 2970 n
592 3165 m
583 3162 n
577 3153 n
574 3138 n
574 3129 n
577 3114 n
583 3105 n
592 3102 n
598 3102 n
607 3105 n
613 3114 n
616 3129 n
616 3138 n
613 3153 n
607 3162 n
598 3165 n
592 3165 n
592 3165 m
586 3162 n
583 3159 n
580 3153 n
577 3138 n
577 3129 n
580 3114 n
583 3108 n
586 3105 n
592 3102 n
598 3102 m
604 3105 n
607 3108 n
610 3114 n
613 3129 n
613 3138 n
610 3153 n
607 3159 n
604 3162 n
598 3165 n
661 3132 m
640 3108 m
637 3105 n
640 3102 n
643 3105 n
640 3108 n
691 3132 m
673 3153 m
679 3156 n
688 3165 n
688 3102 n
685 3162 m
685 3102 n
673 3102 m
700 3102 n
586 3360 m
577 3357 n
571 3348 n
568 3333 n
568 3324 n
571 3309 n
577 3300 n
586 3297 n
592 3297 n
601 3300 n
607 3309 n
610 3324 n
610 3333 n
607 3348 n
601 3357 n
592 3360 n
586 3360 n
586 3360 m
580 3357 n
577 3354 n
574 3348 n
571 3333 n
571 3324 n
574 3309 n
577 3303 n
580 3300 n
586 3297 n
592 3297 m
598 3300 n
601 3303 n
604 3309 n
607 3324 n
607 3333 n
604 3348 n
601 3354 n
598 3357 n
592 3360 n
655 3327 m
634 3303 m
631 3300 n
634 3297 n
637 3300 n
634 3303 n
685 3327 m
661 3348 m
664 3345 n
661 3342 n
658 3345 n
658 3348 n
661 3354 n
664 3357 n
673 3360 n
685 3360 n
694 3357 n
697 3354 n
700 3348 n
700 3342 n
697 3336 n
688 3330 n
673 3324 n
667 3321 n
661 3315 n
658 3306 n
658 3297 n
685 3360 m
691 3357 n
694 3354 n
697 3348 n
697 3342 n
694 3336 n
685 3330 n
673 3324 n
658 3303 m
661 3306 n
667 3306 n
682 3300 n
691 3300 n
697 3303 n
700 3306 n
667 3306 m
682 3297 n
694 3297 n
697 3300 n
700 3306 n
700 3312 n
407 3150 m
470 3126 n
407 3150 m
470 3174 n
416 3150 m
470 3171 n
467 3129 m
467 3171 n
470 3126 m
470 3174 n
440 3216 m
407 3195 m
470 3195 n
407 3198 m
470 3198 n
425 3216 m
449 3216 n
407 3186 m
407 3234 n
425 3234 n
407 3231 n
437 3198 m
437 3216 n
470 3186 m
470 3234 n
452 3234 n
470 3231 n
0 0 4096 4096 s
/solid f
/solid f
883 2570 p
883 2570 p
883 2570 p
883 2570 p
883 2570 p
883 2570 p
883 2570 p
883 2570 p
883 2570 p
883 2570 p
883 2570 p
883 2570 p
883 2570 p
883 2570 p
883 2570 p
883 2570 p
883 2570 p
883 2570 p
883 2570 p
883 2570 p
883 2570 p
883 2570 p
883 2570 p
883 2570 p
883 2570 p
883 2570 p
883 2570 p
883 2570 p
883 2570 p
883 2570 p
883 2570 p
883 2570 p
883 2570 p
883 2570 p
883 2570 p
883 2570 p
883 2570 p
883 2570 p
883 2570 p
883 2570 p
883 2570 p
883 2570 p
883 2570 p
883 2570 p
883 2570 p
883 2570 p
883 2570 p
883 2570 p
883 2570 p
883 2570 p
883 2570 p
883 2570 p
883 2570 p
883 2570 p
883 2570 p
883 2570 p
883 2570 p
883 2570 p
883 2570 p
883 2570 p
883 2570 p
883 2570 p
883 2570 p
883 2570 p
883 2570 p
883 2570 p
883 2570 p
883 2570 p
883 2570 p
883 2570 p
883 2570 p
883 2570 p
883 2570 p
883 2570 p
883 2570 p
883 2570 p
883 2570 p
883 2570 p
883 2570 p
883 2570 p
883 2570 p
883 2570 p
883 2570 p
883 2570 p
883 2570 p
883 2570 p
883 2570 p
883 2570 p
883 2570 p
883 2570 p
883 2570 p
883 2570 p
883 2570 p
883 2570 p
883 2570 p
883 2570 p
883 2570 p
883 2570 p
883 2570 p
883 2570 p
883 2570 p
883 2570 p
883 2570 p
883 2570 p
883 2570 p
883 2570 p
883 2570 p
883 2570 p
883 2570 p
883 2570 p
883 2570 p
883 2570 p
883 2570 p
883 2570 p
883 2570 p
883 2570 p
883 2570 p
883 2570 p
883 2570 p
883 2570 p
883 2570 p
883 2570 p
883 2570 p
883 2570 p
883 2570 p
883 2570 p
883 2570 p
883 2570 p
883 2570 p
883 2570 p
883 2570 p
883 2570 p
883 2570 p
883 2570 p
883 2570 p
883 2570 p
883 2570 p
883 2570 p
883 2570 p
883 2570 p
883 2570 p
883 2570 p
883 2570 p
883 2570 p
883 2570 p
883 2570 p
883 2570 p
883 2570 p
883 2570 p
883 2570 p
883 2570 p
883 2570 p
883 2570 p
883 2570 p
883 2570 p
883 2570 p
883 2570 p
883 2570 p
883 2570 p
883 2570 p
883 2570 p
883 2570 p
883 2570 p
883 2570 p
/shortdashed f
883 2570 m
1016 2632 n
1149 2642 n
1282 2608 n
1415 2566 n
1548 2530 n
1681 2507 n
1814 2495 n
1947 2491 n
2080 2493 n
2212 2498 n
2345 2505 n
2478 2513 n
2611 2521 n
2744 2529 n
2877 2537 n
3010 2544 n
3143 2552 n
3276 2558 n
3409 2565 n
3542 2571 n
/solid f
883 2570 m
1016 2647 n
1149 2708 n
1282 2746 n
1415 2772 n
1548 2787 n
1681 2797 n
1814 2803 n
1947 2809 n
2080 2815 n
2212 2821 n
2345 2827 n
2478 2833 n
2611 2838 n
2744 2843 n
2877 2847 n
3010 2851 n
3143 2855 n
3276 2858 n
3409 2860 n
3542 2863 n
/solid f
750 2450 m
3675 2450 n
3675 2937 n
750 2937 n
750 2450 n
883 2450 m
883 2550 n
883 2837 m
883 2937 n
1149 2450 m
1149 2500 n
1149 2887 m
1149 2937 n
1415 2450 m
1415 2500 n
1415 2887 m
1415 2937 n
1681 2450 m
1681 2500 n
1681 2887 m
1681 2937 n
1947 2450 m
1947 2500 n
1947 2887 m
1947 2937 n
2212 2450 m
2212 2550 n
2212 2837 m
2212 2937 n
2478 2450 m
2478 2500 n
2478 2887 m
2478 2937 n
2744 2450 m
2744 2500 n
2744 2887 m
2744 2937 n
3010 2450 m
3010 2500 n
3010 2887 m
3010 2937 n
3276 2450 m
3276 2500 n
3276 2887 m
3276 2937 n
3542 2450 m
3542 2550 n
3542 2837 m
3542 2937 n
750 2463 m
800 2463 n
3625 2463 m
3675 2463 n
750 2490 m
800 2490 n
3625 2490 m
3675 2490 n
750 2517 m
800 2517 n
3625 2517 m
3675 2517 n
750 2543 m
800 2543 n
3625 2543 m
3675 2543 n
750 2570 m
850 2570 n
3575 2570 m
3675 2570 n
750 2597 m
800 2597 n
3625 2597 m
3675 2597 n
750 2624 m
800 2624 n
3625 2624 m
3675 2624 n
750 2650 m
800 2650 n
3625 2650 m
3675 2650 n
750 2677 m
800 2677 n
3625 2677 m
3675 2677 n
750 2704 m
850 2704 n
3575 2704 m
3675 2704 n
750 2730 m
800 2730 n
3625 2730 m
3675 2730 n
750 2757 m
800 2757 n
3625 2757 m
3675 2757 n
750 2784 m
800 2784 n
3625 2784 m
3675 2784 n
750 2811 m
800 2811 n
3625 2811 m
3675 2811 n
750 2837 m
850 2837 n
3575 2837 m
3675 2837 n
750 2864 m
800 2864 n
3625 2864 m
3675 2864 n
750 2891 m
800 2891 n
3625 2891 m
3675 2891 n
750 2917 m
800 2917 n
3625 2917 m
3675 2917 n
943 2908 m
937 2902 n
931 2893 n
925 2881 n
922 2866 n
922 2854 n
925 2839 n
931 2827 n
937 2818 n
943 2812 n
937 2902 m
931 2890 n
928 2881 n
925 2866 n
925 2854 n
928 2839 n
931 2830 n
937 2818 n
988 2863 m
997 2866 m
994 2863 n
997 2860 n
1000 2863 n
1000 2866 n
994 2872 n
988 2875 n
979 2875 n
970 2872 n
964 2866 n
961 2857 n
961 2851 n
964 2842 n
970 2836 n
979 2833 n
985 2833 n
994 2836 n
1000 2842 n
979 2875 m
973 2872 n
967 2866 n
964 2857 n
964 2851 n
967 2842 n
973 2836 n
979 2833 n
1045 2863 m
1018 2908 m
1024 2902 n
1030 2893 n
1036 2881 n
1039 2866 n
1039 2854 n
1036 2839 n
1030 2827 n
1024 2818 n
1018 2812 n
1024 2902 m
1030 2890 n
1033 2881 n
1036 2866 n
1036 2854 n
1033 2839 n
1030 2830 n
1024 2818 n
676 2603 m
667 2600 n
661 2591 n
658 2576 n
658 2567 n
661 2552 n
667 2543 n
676 2540 n
682 2540 n
691 2543 n
697 2552 n
700 2567 n
700 2576 n
697 2591 n
691 2600 n
682 2603 n
676 2603 n
676 2603 m
670 2600 n
667 2597 n
664 2591 n
661 2576 n
661 2567 n
664 2552 n
667 2546 n
670 2543 n
676 2540 n
682 2540 m
688 2543 n
691 2546 n
694 2552 n
697 2567 n
697 2576 n
694 2591 n
691 2597 n
688 2600 n
682 2603 n
586 2737 m
577 2734 n
571 2725 n
568 2710 n
568 2701 n
571 2686 n
577 2677 n
586 2674 n
592 2674 n
601 2677 n
607 2686 n
610 2701 n
610 2710 n
607 2725 n
601 2734 n
592 2737 n
586 2737 n
586 2737 m
580 2734 n
577 2731 n
574 2725 n
571 2710 n
571 2701 n
574 2686 n
577 2680 n
580 2677 n
586 2674 n
592 2674 m
598 2677 n
601 2680 n
604 2686 n
607 2701 n
607 2710 n
604 2725 n
601 2731 n
598 2734 n
592 2737 n
655 2704 m
634 2680 m
631 2677 n
634 2674 n
637 2677 n
634 2680 n
685 2704 m
661 2725 m
664 2722 n
661 2719 n
658 2722 n
658 2725 n
661 2731 n
664 2734 n
673 2737 n
685 2737 n
694 2734 n
697 2731 n
700 2725 n
700 2719 n
697 2713 n
688 2707 n
673 2701 n
667 2698 n
661 2692 n
658 2683 n
658 2674 n
685 2737 m
691 2734 n
694 2731 n
697 2725 n
697 2719 n
694 2713 n
685 2707 n
673 2701 n
658 2680 m
661 2683 n
667 2683 n
682 2677 n
691 2677 n
697 2680 n
700 2683 n
667 2683 m
682 2674 n
694 2674 n
697 2677 n
700 2683 n
700 2689 n
583 2870 m
574 2867 n
568 2858 n
565 2843 n
565 2834 n
568 2819 n
574 2810 n
583 2807 n
589 2807 n
598 2810 n
604 2819 n
607 2834 n
607 2843 n
604 2858 n
598 2867 n
589 2870 n
583 2870 n
583 2870 m
577 2867 n
574 2864 n
571 2858 n
568 2843 n
568 2834 n
571 2819 n
574 2813 n
577 2810 n
583 2807 n
589 2807 m
595 2810 n
598 2813 n
601 2819 n
604 2834 n
604 2843 n
601 2858 n
598 2864 n
595 2867 n
589 2870 n
652 2837 m
631 2813 m
628 2810 n
631 2807 n
634 2810 n
631 2813 n
682 2837 m
682 2864 m
682 2807 n
685 2870 m
685 2807 n
685 2870 m
652 2825 n
700 2825 n
673 2807 m
694 2807 n
407 2663 m
470 2639 n
407 2663 m
470 2687 n
416 2663 m
470 2684 n
467 2642 m
467 2684 n
470 2639 m
470 2687 n
440 2729 m
407 2708 m
470 2708 n
407 2711 m
470 2711 n
425 2729 m
449 2729 n
407 2699 m
407 2747 n
425 2747 n
407 2744 n
437 2711 m
437 2729 n
470 2699 m
470 2747 n
452 2747 n
470 2744 n
0 0 4096 4096 s
/solid f
/solid f
883 1962 p
883 1962 p
883 1962 p
883 1962 p
883 1962 p
883 1962 p
883 1962 p
883 1962 p
883 1962 p
883 1962 p
883 1962 p
883 1962 p
883 1962 p
883 1962 p
883 1962 p
883 1962 p
883 1962 p
883 1962 p
883 1962 p
883 1962 p
883 1962 p
883 1962 p
883 1962 p
883 1962 p
883 1962 p
883 1962 p
883 1962 p
883 1962 p
883 1962 p
883 1962 p
883 1962 p
883 1962 p
883 1962 p
883 1962 p
883 1962 p
883 1962 p
883 1962 p
883 1962 p
883 1962 p
883 1962 p
883 1962 p
883 1962 p
883 1962 p
883 1962 p
883 1962 p
883 1962 p
883 1962 p
883 1962 p
883 1962 p
883 1962 p
883 1962 p
883 1962 p
883 1962 p
883 1962 p
883 1962 p
883 1962 p
883 1962 p
883 1962 p
883 1962 p
883 1962 p
883 1962 p
883 1962 p
883 1962 p
883 1962 p
883 1962 p
883 1962 p
883 1962 p
883 1962 p
883 1962 p
883 1962 p
883 1962 p
883 1962 p
883 1962 p
883 1962 p
883 1962 p
883 1962 p
883 1962 p
883 1962 p
883 1962 p
883 1962 p
883 1962 p
883 1962 p
883 1962 p
883 1962 p
883 1962 p
883 1962 p
883 1962 p
883 1962 p
883 1962 p
883 1962 p
883 1962 p
883 1962 p
883 1962 p
883 1962 p
883 1962 p
883 1962 p
883 1962 p
883 1962 p
883 1962 p
883 1962 p
883 1962 p
883 1962 p
883 1962 p
883 1962 p
883 1962 p
883 1962 p
883 1962 p
883 1962 p
883 1962 p
883 1962 p
883 1962 p
883 1962 p
883 1962 p
883 1962 p
883 1962 p
883 1962 p
883 1962 p
883 1962 p
883 1962 p
883 1962 p
883 1962 p
883 1962 p
883 1962 p
883 1962 p
883 1962 p
883 1962 p
883 1962 p
883 1962 p
883 1962 p
883 1962 p
883 1962 p
883 1962 p
883 1962 p
883 1962 p
883 1962 p
883 1962 p
883 1962 p
883 1962 p
883 1962 p
883 1962 p
883 1962 p
883 1962 p
883 1962 p
883 1962 p
883 1962 p
883 1962 p
883 1962 p
883 1962 p
883 1962 p
883 1962 p
883 1962 p
883 1962 p
883 1962 p
883 1962 p
883 1962 p
883 1962 p
883 1962 p
883 1962 p
883 1962 p
883 1962 p
883 1962 p
883 1962 p
883 1962 p
883 1962 p
883 1962 p
883 1962 p
883 1962 p
883 1962 p
883 1962 p
883 1962 p
883 1962 p
883 1962 p
883 1962 p
883 1962 p
883 1962 p
883 1962 p
883 1962 p
883 1962 p
883 1962 p
883 1962 p
883 1962 p
883 1962 p
883 1962 p
883 1962 p
883 1962 p
883 1962 p
883 1962 p
883 1962 p
883 1962 p
883 1962 p
883 1962 p
883 1962 p
883 1962 p
883 1962 p
883 1962 p
883 1962 p
883 1962 p
883 1962 p
883 1962 p
883 1962 p
883 1962 p
883 1962 p
883 1962 p
883 1962 p
883 1962 p
883 1962 p
/shortdashed f
883 1962 m
1016 2136 n
1149 2273 n
1282 2367 n
1415 2412 n
1548 2401 n
1681 2315 n
1814 2242 n
1947 2190 n
2080 2158 n
2212 2141 n
2345 2136 n
2478 2138 n
2611 2146 n
2744 2157 n
2877 2170 n
3010 2184 n
3143 2198 n
3276 2213 n
3409 2227 n
3542 2241 n
/solid f
883 1962 m
1016 2136 n
1149 2273 n
1282 2367 n
1415 2412 n
1548 2401 n
1681 2315 n
1814 2242 n
1947 2190 n
2080 2158 n
2212 2141 n
2345 2136 n
2478 2138 n
2611 2146 n
2744 2157 n
2877 2170 n
3010 2184 n
3143 2198 n
3276 2213 n
3409 2227 n
3542 2241 n
/solid f
750 1962 m
3675 1962 n
3675 2449 n
750 2449 n
750 1962 n
883 1962 m
883 2062 n
883 2349 m
883 2449 n
1149 1962 m
1149 2012 n
1149 2399 m
1149 2449 n
1415 1962 m
1415 2012 n
1415 2399 m
1415 2449 n
1681 1962 m
1681 2012 n
1681 2399 m
1681 2449 n
1947 1962 m
1947 2012 n
1947 2399 m
1947 2449 n
2212 1962 m
2212 2062 n
2212 2349 m
2212 2449 n
2478 1962 m
2478 2012 n
2478 2399 m
2478 2449 n
2744 1962 m
2744 2012 n
2744 2399 m
2744 2449 n
3010 1962 m
3010 2012 n
3010 2399 m
3010 2449 n
3276 1962 m
3276 2012 n
3276 2399 m
3276 2449 n
3542 1962 m
3542 2062 n
3542 2349 m
3542 2449 n
750 2001 m
800 2001 n
3625 2001 m
3675 2001 n
750 2040 m
800 2040 n
3625 2040 m
3675 2040 n
750 2079 m
800 2079 n
3625 2079 m
3675 2079 n
750 2118 m
800 2118 n
3625 2118 m
3675 2118 n
750 2157 m
850 2157 n
3575 2157 m
3675 2157 n
750 2196 m
800 2196 n
3625 2196 m
3675 2196 n
750 2235 m
800 2235 n
3625 2235 m
3675 2235 n
750 2274 m
800 2274 n
3625 2274 m
3675 2274 n
750 2313 m
800 2313 n
3625 2313 m
3675 2313 n
750 2352 m
850 2352 n
3575 2352 m
3675 2352 n
750 2391 m
800 2391 n
3625 2391 m
3675 2391 n
750 2430 m
800 2430 n
3625 2430 m
3675 2430 n
940 2420 m
934 2414 n
928 2405 n
922 2393 n
919 2378 n
919 2366 n
922 2351 n
928 2339 n
934 2330 n
940 2324 n
934 2414 m
928 2402 n
925 2393 n
922 2378 n
922 2366 n
925 2351 n
928 2342 n
934 2330 n
985 2375 m
994 2408 m
994 2345 n
997 2408 m
997 2345 n
994 2378 m
988 2384 n
982 2387 n
976 2387 n
967 2384 n
961 2378 n
958 2369 n
958 2363 n
961 2354 n
967 2348 n
976 2345 n
982 2345 n
988 2348 n
994 2354 n
976 2387 m
970 2384 n
964 2378 n
961 2369 n
961 2363 n
964 2354 n
970 2348 n
976 2345 n
985 2408 m
997 2408 n
994 2345 m
1006 2345 n
1048 2375 m
1021 2420 m
1027 2414 n
1033 2405 n
1039 2393 n
1042 2378 n
1042 2366 n
1039 2351 n
1033 2339 n
1027 2330 n
1021 2324 n
1027 2414 m
1033 2402 n
1036 2393 n
1039 2378 n
1039 2366 n
1036 2351 n
1033 2342 n
1027 2330 n
676 1995 m
667 1992 n
661 1983 n
658 1968 n
658 1959 n
661 1944 n
667 1935 n
676 1932 n
682 1932 n
691 1935 n
697 1944 n
700 1959 n
700 1968 n
697 1983 n
691 1992 n
682 1995 n
676 1995 n
676 1995 m
670 1992 n
667 1989 n
664 1983 n
661 1968 n
661 1959 n
664 1944 n
667 1938 n
670 1935 n
676 1932 n
682 1932 m
688 1935 n
691 1938 n
694 1944 n
697 1959 n
697 1968 n
694 1983 n
691 1989 n
688 1992 n
682 1995 n
592 2190 m
583 2187 n
577 2178 n
574 2163 n
574 2154 n
577 2139 n
583 2130 n
592 2127 n
598 2127 n
607 2130 n
613 2139 n
616 2154 n
616 2163 n
613 2178 n
607 2187 n
598 2190 n
592 2190 n
592 2190 m
586 2187 n
583 2184 n
580 2178 n
577 2163 n
577 2154 n
580 2139 n
583 2133 n
586 2130 n
592 2127 n
598 2127 m
604 2130 n
607 2133 n
610 2139 n
613 2154 n
613 2163 n
610 2178 n
607 2184 n
604 2187 n
598 2190 n
661 2157 m
640 2133 m
637 2130 n
640 2127 n
643 2130 n
640 2133 n
691 2157 m
673 2178 m
679 2181 n
688 2190 n
688 2127 n
685 2187 m
685 2127 n
673 2127 m
700 2127 n
586 2385 m
577 2382 n
571 2373 n
568 2358 n
568 2349 n
571 2334 n
577 2325 n
586 2322 n
592 2322 n
601 2325 n
607 2334 n
610 2349 n
610 2358 n
607 2373 n
601 2382 n
592 2385 n
586 2385 n
586 2385 m
580 2382 n
577 2379 n
574 2373 n
571 2358 n
571 2349 n
574 2334 n
577 2328 n
580 2325 n
586 2322 n
592 2322 m
598 2325 n
601 2328 n
604 2334 n
607 2349 n
607 2358 n
604 2373 n
601 2379 n
598 2382 n
592 2385 n
655 2352 m
634 2328 m
631 2325 n
634 2322 n
637 2325 n
634 2328 n
685 2352 m
661 2373 m
664 2370 n
661 2367 n
658 2370 n
658 2373 n
661 2379 n
664 2382 n
673 2385 n
685 2385 n
694 2382 n
697 2379 n
700 2373 n
700 2367 n
697 2361 n
688 2355 n
673 2349 n
667 2346 n
661 2340 n
658 2331 n
658 2322 n
685 2385 m
691 2382 n
694 2379 n
697 2373 n
697 2367 n
694 2361 n
685 2355 n
673 2349 n
658 2328 m
661 2331 n
667 2331 n
682 2325 n
691 2325 n
697 2328 n
700 2331 n
667 2331 m
682 2322 n
694 2322 n
697 2325 n
700 2331 n
700 2337 n
407 2175 m
470 2151 n
407 2175 m
470 2199 n
416 2175 m
470 2196 n
467 2154 m
467 2196 n
470 2151 m
470 2199 n
440 2241 m
407 2220 m
470 2220 n
407 2223 m
470 2223 n
425 2241 m
449 2241 n
407 2211 m
407 2259 n
425 2259 n
407 2256 n
437 2223 m
437 2241 n
470 2211 m
470 2259 n
452 2259 n
470 2256 n
0 0 4096 4096 s
/solid f
/solid f
883 1588 p
883 1588 p
883 1588 p
883 1588 p
883 1588 p
883 1588 p
883 1588 p
883 1588 p
883 1588 p
883 1588 p
883 1588 p
883 1588 p
883 1588 p
883 1588 p
883 1588 p
883 1588 p
883 1588 p
883 1588 p
883 1588 p
883 1588 p
883 1588 p
883 1588 p
883 1588 p
883 1588 p
883 1588 p
883 1588 p
883 1588 p
883 1588 p
883 1588 p
883 1588 p
883 1588 p
883 1588 p
883 1588 p
883 1588 p
883 1588 p
883 1588 p
883 1588 p
883 1588 p
883 1588 p
883 1588 p
883 1588 p
883 1588 p
883 1588 p
883 1588 p
883 1588 p
883 1588 p
883 1588 p
883 1588 p
883 1588 p
883 1588 p
883 1588 p
883 1588 p
883 1588 p
883 1588 p
883 1588 p
883 1588 p
883 1588 p
883 1588 p
883 1588 p
883 1588 p
883 1588 p
883 1588 p
883 1588 p
883 1588 p
883 1588 p
883 1588 p
883 1588 p
883 1588 p
883 1588 p
883 1588 p
883 1588 p
883 1588 p
883 1588 p
883 1588 p
883 1588 p
883 1588 p
883 1588 p
883 1588 p
883 1588 p
883 1588 p
883 1588 p
883 1588 p
883 1588 p
883 1588 p
883 1588 p
883 1588 p
883 1588 p
883 1588 p
883 1588 p
883 1588 p
883 1588 p
883 1588 p
883 1588 p
883 1588 p
883 1588 p
883 1588 p
883 1588 p
883 1588 p
883 1588 p
883 1588 p
883 1588 p
883 1588 p
883 1588 p
883 1588 p
883 1588 p
883 1588 p
883 1588 p
883 1588 p
883 1588 p
883 1588 p
883 1588 p
883 1588 p
883 1588 p
883 1588 p
883 1588 p
883 1588 p
883 1588 p
883 1588 p
883 1588 p
883 1588 p
883 1588 p
883 1588 p
883 1588 p
883 1588 p
883 1588 p
883 1588 p
883 1588 p
883 1588 p
883 1588 p
883 1588 p
883 1588 p
883 1588 p
883 1588 p
883 1588 p
883 1588 p
883 1588 p
883 1588 p
883 1588 p
883 1588 p
883 1588 p
883 1588 p
883 1588 p
883 1588 p
883 1588 p
883 1588 p
883 1588 p
883 1588 p
883 1588 p
883 1588 p
883 1588 p
883 1588 p
883 1588 p
883 1588 p
883 1588 p
883 1588 p
883 1588 p
883 1588 p
883 1588 p
883 1588 p
883 1588 p
883 1588 p
883 1588 p
883 1588 p
883 1588 p
883 1588 p
883 1588 p
883 1588 p
883 1588 p
883 1588 p
883 1588 p
883 1588 p
883 1588 p
883 1588 p
883 1588 p
883 1588 p
883 1588 p
883 1588 p
883 1588 p
883 1588 p
883 1588 p
883 1588 p
883 1588 p
883 1588 p
883 1588 p
883 1588 p
883 1588 p
883 1588 p
883 1588 p
883 1588 p
883 1588 p
883 1588 p
883 1588 p
883 1588 p
883 1588 p
883 1588 p
883 1588 p
883 1588 p
883 1588 p
883 1588 p
883 1588 p
883 1588 p
883 1588 p
883 1588 p
883 1588 p
883 1588 p
883 1588 p
883 1588 p
883 1588 p
883 1588 p
883 1588 p
883 1588 p
883 1588 p
883 1588 p
883 1588 p
883 1588 p
883 1588 p
883 1588 p
883 1588 p
883 1588 p
883 1588 p
883 1588 p
883 1588 p
883 1588 p
883 1588 p
883 1588 p
883 1588 p
883 1588 p
883 1588 p
883 1588 p
883 1588 p
883 1588 p
883 1588 p
883 1588 p
883 1588 p
883 1588 p
883 1588 p
883 1588 p
883 1588 p
883 1588 p
883 1588 p
883 1588 p
883 1588 p
883 1588 p
883 1588 p
883 1588 p
883 1588 p
883 1588 p
883 1588 p
/shortdashed f
883 1588 m
1016 1563 n
1149 1514 n
1282 1493 n
1415 1499 n
1548 1523 n
1681 1555 n
1814 1590 n
1947 1624 n
2080 1658 n
2212 1691 n
2345 1726 n
2478 1757 n
2611 1786 n
2744 1813 n
2877 1839 n
3010 1864 n
3143 1887 n
3276 1909 n
3409 1930 n
3542 1949 n
/solid f
883 1588 m
1016 1564 n
1149 1515 n
1282 1493 n
1415 1499 n
1548 1523 n
1681 1555 n
1814 1590 n
1947 1625 n
2080 1659 n
2212 1692 n
2345 1724 n
2478 1755 n
2611 1784 n
2744 1812 n
2877 1839 n
3010 1864 n
3143 1887 n
3276 1909 n
3409 1929 n
3542 1948 n
/solid f
750 1475 m
3675 1475 n
3675 1962 n
750 1962 n
750 1475 n
883 1475 m
883 1575 n
883 1862 m
883 1962 n
1149 1475 m
1149 1525 n
1149 1912 m
1149 1962 n
1415 1475 m
1415 1525 n
1415 1912 m
1415 1962 n
1681 1475 m
1681 1525 n
1681 1912 m
1681 1962 n
1947 1475 m
1947 1525 n
1947 1912 m
1947 1962 n
2212 1475 m
2212 1575 n
2212 1862 m
2212 1962 n
2478 1475 m
2478 1525 n
2478 1912 m
2478 1962 n
2744 1475 m
2744 1525 n
2744 1912 m
2744 1962 n
3010 1475 m
3010 1525 n
3010 1912 m
3010 1962 n
3276 1475 m
3276 1525 n
3276 1912 m
3276 1962 n
3542 1475 m
3542 1575 n
3542 1862 m
3542 1962 n
750 1497 m
800 1497 n
3625 1497 m
3675 1497 n
750 1527 m
800 1527 n
3625 1527 m
3675 1527 n
750 1557 m
800 1557 n
3625 1557 m
3675 1557 n
750 1588 m
850 1588 n
3575 1588 m
3675 1588 n
750 1618 m
800 1618 n
3625 1618 m
3675 1618 n
750 1648 m
800 1648 n
3625 1648 m
3675 1648 n
750 1678 m
800 1678 n
3625 1678 m
3675 1678 n
750 1708 m
800 1708 n
3625 1708 m
3675 1708 n
750 1738 m
850 1738 n
3575 1738 m
3675 1738 n
750 1768 m
800 1768 n
3625 1768 m
3675 1768 n
750 1798 m
800 1798 n
3625 1798 m
3675 1798 n
750 1828 m
800 1828 n
3625 1828 m
3675 1828 n
750 1858 m
800 1858 n
3625 1858 m
3675 1858 n
750 1888 m
850 1888 n
3575 1888 m
3675 1888 n
750 1918 m
800 1918 n
3625 1918 m
3675 1918 n
750 1948 m
800 1948 n
3625 1948 m
3675 1948 n
943 1933 m
937 1927 n
931 1918 n
925 1906 n
922 1891 n
922 1879 n
925 1864 n
931 1852 n
937 1843 n
943 1837 n
937 1927 m
931 1915 n
928 1906 n
925 1891 n
925 1879 n
928 1864 n
931 1855 n
937 1843 n
988 1888 m
964 1882 m
1000 1882 n
1000 1888 n
997 1894 n
994 1897 n
988 1900 n
979 1900 n
970 1897 n
964 1891 n
961 1882 n
961 1876 n
964 1867 n
970 1861 n
979 1858 n
985 1858 n
994 1861 n
1000 1867 n
997 1882 m
997 1891 n
994 1897 n
979 1900 m
973 1897 n
967 1891 n
964 1882 n
964 1876 n
967 1867 n
973 1861 n
979 1858 n
1045 1888 m
1018 1933 m
1024 1927 n
1030 1918 n
1036 1906 n
1039 1891 n
1039 1879 n
1036 1864 n
1030 1852 n
1024 1843 n
1018 1837 n
1024 1927 m
1030 1915 n
1033 1906 n
1036 1891 n
1036 1879 n
1033 1864 n
1030 1855 n
1024 1843 n
676 1621 m
667 1618 n
661 1609 n
658 1594 n
658 1585 n
661 1570 n
667 1561 n
676 1558 n
682 1558 n
691 1561 n
697 1570 n
700 1585 n
700 1594 n
697 1609 n
691 1618 n
682 1621 n
676 1621 n
676 1621 m
670 1618 n
667 1615 n
664 1609 n
661 1594 n
661 1585 n
664 1570 n
667 1564 n
670 1561 n
676 1558 n
682 1558 m
688 1561 n
691 1564 n
694 1570 n
697 1585 n
697 1594 n
694 1609 n
691 1615 n
688 1618 n
682 1621 n
523 1771 m
514 1768 n
508 1759 n
505 1744 n
505 1735 n
508 1720 n
514 1711 n
523 1708 n
529 1708 n
538 1711 n
544 1720 n
547 1735 n
547 1744 n
544 1759 n
538 1768 n
529 1771 n
523 1771 n
523 1771 m
517 1768 n
514 1765 n
511 1759 n
508 1744 n
508 1735 n
511 1720 n
514 1714 n
517 1711 n
523 1708 n
529 1708 m
535 1711 n
538 1714 n
541 1720 n
544 1735 n
544 1744 n
541 1759 n
538 1765 n
535 1768 n
529 1771 n
592 1738 m
571 1714 m
568 1711 n
571 1708 n
574 1711 n
571 1714 n
622 1738 m
613 1771 m
604 1768 n
598 1759 n
595 1744 n
595 1735 n
598 1720 n
604 1711 n
613 1708 n
619 1708 n
628 1711 n
634 1720 n
637 1735 n
637 1744 n
634 1759 n
628 1768 n
619 1771 n
613 1771 n
613 1771 m
607 1768 n
604 1765 n
601 1759 n
598 1744 n
598 1735 n
601 1720 n
604 1714 n
607 1711 n
613 1708 n
619 1708 m
625 1711 n
628 1714 n
631 1720 n
634 1735 n
634 1744 n
631 1759 n
628 1765 n
625 1768 n
619 1771 n
682 1738 m
682 1765 m
682 1708 n
685 1771 m
685 1708 n
685 1771 m
652 1726 n
700 1726 n
673 1708 m
694 1708 n
526 1921 m
517 1918 n
511 1909 n
508 1894 n
508 1885 n
511 1870 n
517 1861 n
526 1858 n
532 1858 n
541 1861 n
547 1870 n
550 1885 n
550 1894 n
547 1909 n
541 1918 n
532 1921 n
526 1921 n
526 1921 m
520 1918 n
517 1915 n
514 1909 n
511 1894 n
511 1885 n
514 1870 n
517 1864 n
520 1861 n
526 1858 n
532 1858 m
538 1861 n
541 1864 n
544 1870 n
547 1885 n
547 1894 n
544 1909 n
541 1915 n
538 1918 n
532 1921 n
595 1888 m
574 1864 m
571 1861 n
574 1858 n
577 1861 n
574 1864 n
625 1888 m
616 1921 m
607 1918 n
601 1909 n
598 1894 n
598 1885 n
601 1870 n
607 1861 n
616 1858 n
622 1858 n
631 1861 n
637 1870 n
640 1885 n
640 1894 n
637 1909 n
631 1918 n
622 1921 n
616 1921 n
616 1921 m
610 1918 n
607 1915 n
604 1909 n
601 1894 n
601 1885 n
604 1870 n
607 1864 n
610 1861 n
616 1858 n
622 1858 m
628 1861 n
631 1864 n
634 1870 n
637 1885 n
637 1894 n
634 1909 n
631 1915 n
628 1918 n
622 1921 n
685 1888 m
673 1921 m
664 1918 n
661 1912 n
661 1903 n
664 1897 n
673 1894 n
685 1894 n
694 1897 n
697 1903 n
697 1912 n
694 1918 n
685 1921 n
673 1921 n
673 1921 m
667 1918 n
664 1912 n
664 1903 n
667 1897 n
673 1894 n
685 1894 m
691 1897 n
694 1903 n
694 1912 n
691 1918 n
685 1921 n
673 1894 m
664 1891 n
661 1888 n
658 1882 n
658 1870 n
661 1864 n
664 1861 n
673 1858 n
685 1858 n
694 1861 n
697 1864 n
700 1870 n
700 1882 n
697 1888 n
694 1891 n
685 1894 n
673 1894 m
667 1891 n
664 1888 n
661 1882 n
661 1870 n
664 1864 n
667 1861 n
673 1858 n
685 1858 m
691 1861 n
694 1864 n
697 1870 n
697 1882 n
694 1888 n
691 1891 n
685 1894 n
347 1688 m
410 1664 n
347 1688 m
410 1712 n
356 1688 m
410 1709 n
407 1667 m
407 1709 n
410 1664 m
410 1712 n
380 1754 m
347 1733 m
410 1733 n
347 1736 m
410 1736 n
365 1754 m
389 1754 n
347 1724 m
347 1772 n
365 1772 n
347 1769 n
377 1736 m
377 1754 n
410 1724 m
410 1772 n
392 1772 n
410 1769 n
0 0 4096 4096 s
/solid f
/solid f
883 1185 p
883 1185 p
883 1185 p
883 1185 p
883 1185 p
883 1185 p
883 1185 p
883 1185 p
883 1185 p
883 1185 p
883 1185 p
883 1185 p
883 1185 p
883 1185 p
883 1185 p
883 1185 p
883 1185 p
883 1185 p
883 1185 p
883 1185 p
883 1185 p
883 1185 p
883 1185 p
883 1185 p
883 1185 p
883 1185 p
883 1185 p
883 1185 p
883 1185 p
883 1185 p
883 1185 p
883 1185 p
883 1185 p
883 1185 p
883 1185 p
883 1185 p
883 1185 p
883 1185 p
883 1185 p
883 1185 p
883 1185 p
883 1185 p
883 1185 p
883 1185 p
883 1185 p
883 1185 p
883 1185 p
883 1185 p
883 1185 p
883 1185 p
883 1185 p
883 1185 p
883 1185 p
883 1185 p
883 1185 p
883 1185 p
883 1185 p
883 1185 p
883 1185 p
883 1185 p
883 1185 p
883 1185 p
883 1185 p
883 1185 p
883 1185 p
883 1185 p
883 1185 p
883 1185 p
883 1185 p
883 1185 p
883 1185 p
883 1185 p
883 1185 p
883 1185 p
883 1185 p
883 1185 p
883 1185 p
883 1185 p
883 1185 p
883 1185 p
883 1185 p
883 1185 p
883 1185 p
883 1185 p
883 1185 p
883 1185 p
883 1185 p
883 1185 p
883 1185 p
883 1185 p
883 1185 p
883 1185 p
883 1185 p
883 1185 p
883 1185 p
883 1185 p
883 1185 p
883 1185 p
883 1185 p
883 1185 p
883 1185 p
883 1185 p
883 1185 p
883 1185 p
883 1185 p
883 1185 p
883 1185 p
883 1185 p
883 1185 p
883 1185 p
883 1185 p
883 1185 p
883 1185 p
883 1185 p
883 1185 p
883 1185 p
883 1185 p
883 1185 p
883 1185 p
883 1185 p
883 1185 p
883 1185 p
883 1185 p
883 1185 p
883 1185 p
883 1185 p
883 1185 p
883 1185 p
883 1185 p
883 1185 p
883 1185 p
883 1185 p
883 1185 p
883 1185 p
883 1185 p
883 1185 p
883 1185 p
883 1185 p
883 1185 p
883 1185 p
883 1185 p
883 1185 p
883 1185 p
883 1185 p
883 1185 p
883 1185 p
883 1185 p
883 1185 p
883 1185 p
883 1185 p
883 1185 p
883 1185 p
883 1185 p
883 1185 p
883 1185 p
883 1185 p
883 1185 p
883 1185 p
883 1185 p
883 1185 p
883 1185 p
883 1185 p
883 1185 p
883 1185 p
883 1185 p
883 1185 p
883 1185 p
883 1185 p
883 1185 p
883 1185 p
883 1185 p
883 1185 p
883 1185 p
883 1185 p
883 1185 p
883 1185 p
883 1185 p
883 1185 p
883 1185 p
883 1185 p
883 1185 p
883 1185 p
883 1185 p
883 1185 p
883 1185 p
883 1185 p
883 1185 p
883 1185 p
883 1185 p
883 1185 p
883 1185 p
883 1185 p
883 1185 p
883 1185 p
883 1185 p
883 1185 p
883 1185 p
883 1185 p
883 1185 p
883 1185 p
883 1185 p
883 1185 p
883 1185 p
883 1185 p
883 1185 p
883 1185 p
883 1185 p
883 1185 p
883 1185 p
883 1185 p
883 1185 p
883 1185 p
883 1185 p
883 1185 p
883 1185 p
883 1185 p
883 1185 p
883 1185 p
883 1185 p
883 1185 p
883 1185 p
883 1185 p
883 1185 p
883 1185 p
883 1185 p
883 1185 p
883 1185 p
883 1185 p
883 1185 p
883 1185 p
883 1185 p
883 1185 p
883 1185 p
883 1185 p
883 1185 p
883 1185 p
883 1185 p
883 1185 p
883 1185 p
883 1185 p
883 1185 p
883 1185 p
883 1185 p
883 1185 p
883 1185 p
883 1185 p
883 1185 p
883 1185 p
883 1185 p
883 1185 p
883 1185 p
883 1185 p
883 1185 p
883 1185 p
883 1185 p
883 1185 p
883 1185 p
883 1185 p
883 1185 p
883 1185 p
883 1185 p
883 1185 p
883 1185 p
883 1185 p
883 1185 p
883 1185 p
883 1185 p
883 1185 p
883 1185 p
883 1185 p
883 1185 p
883 1185 p
883 1185 p
883 1185 p
883 1185 p
883 1185 p
883 1185 p
883 1185 p
883 1185 p
883 1185 p
883 1185 p
883 1185 p
883 1185 p
883 1185 p
883 1185 p
883 1185 p
883 1185 p
883 1185 p
883 1185 p
883 1185 p
/shortdashed f
883 1185 m
1016 1295 n
1149 1435 n
1282 1446 n
1415 1372 n
1548 1286 n
1681 1220 n
1814 1179 n
1947 1157 n
2080 1148 n
2212 1148 n
2345 1153 n
2478 1160 n
2611 1170 n
2744 1180 n
2877 1190 n
3010 1201 n
3143 1211 n
3276 1221 n
3409 1230 n
3542 1239 n
/solid f
883 1185 m
1016 1287 n
1149 1396 n
1282 1366 n
1415 1257 n
1548 1158 n
1681 1094 n
1814 1062 n
1947 1050 n
2080 1051 n
2212 1059 n
2345 1071 n
2478 1084 n
2611 1098 n
2744 1113 n
2877 1127 n
3010 1140 n
3143 1153 n
3276 1166 n
3409 1178 n
3542 1189 n
/solid f
750 987 m
3675 987 n
3675 1474 n
750 1474 n
750 987 n
883 987 m
883 1087 n
883 1374 m
883 1474 n
1149 987 m
1149 1037 n
1149 1424 m
1149 1474 n
1415 987 m
1415 1037 n
1415 1424 m
1415 1474 n
1681 987 m
1681 1037 n
1681 1424 m
1681 1474 n
1947 987 m
1947 1037 n
1947 1424 m
1947 1474 n
2212 987 m
2212 1087 n
2212 1374 m
2212 1474 n
2478 987 m
2478 1037 n
2478 1424 m
2478 1474 n
2744 987 m
2744 1037 n
2744 1424 m
2744 1474 n
3010 987 m
3010 1037 n
3010 1424 m
3010 1474 n
3276 987 m
3276 1037 n
3276 1424 m
3276 1474 n
3542 987 m
3542 1087 n
3542 1374 m
3542 1474 n
750 1000 m
800 1000 n
3625 1000 m
3675 1000 n
750 1027 m
800 1027 n
3625 1027 m
3675 1027 n
750 1053 m
850 1053 n
3575 1053 m
3675 1053 n
750 1079 m
800 1079 n
3625 1079 m
3675 1079 n
750 1106 m
800 1106 n
3625 1106 m
3675 1106 n
750 1132 m
800 1132 n
3625 1132 m
3675 1132 n
750 1158 m
800 1158 n
3625 1158 m
3675 1158 n
750 1185 m
850 1185 n
3575 1185 m
3675 1185 n
750 1211 m
800 1211 n
3625 1211 m
3675 1211 n
750 1237 m
800 1237 n
3625 1237 m
3675 1237 n
750 1264 m
800 1264 n
3625 1264 m
3675 1264 n
750 1290 m
800 1290 n
3625 1290 m
3675 1290 n
750 1316 m
850 1316 n
3575 1316 m
3675 1316 n
750 1343 m
800 1343 n
3625 1343 m
3675 1343 n
750 1369 m
800 1369 n
3625 1369 m
3675 1369 n
750 1395 m
800 1395 n
3625 1395 m
3675 1395 n
750 1422 m
800 1422 n
3625 1422 m
3675 1422 n
750 1448 m
850 1448 n
3575 1448 m
3675 1448 n
952 1445 m
946 1439 n
940 1430 n
934 1418 n
931 1403 n
931 1391 n
934 1376 n
940 1364 n
946 1355 n
952 1349 n
946 1439 m
940 1427 n
937 1418 n
934 1403 n
934 1391 n
937 1376 n
940 1367 n
946 1355 n
997 1400 m
991 1430 m
988 1427 n
991 1424 n
994 1427 n
994 1430 n
991 1433 n
985 1433 n
979 1430 n
976 1424 n
976 1370 n
985 1433 m
982 1430 n
979 1424 n
979 1370 n
967 1412 m
991 1412 n
967 1370 m
988 1370 n
1036 1400 m
1009 1445 m
1015 1439 n
1021 1430 n
1027 1418 n
1030 1403 n
1030 1391 n
1027 1376 n
1021 1364 n
1015 1355 n
1009 1349 n
1015 1439 m
1021 1427 n
1024 1418 n
1027 1403 n
1027 1391 n
1024 1376 n
1021 1367 n
1015 1355 n
499 1050 m
553 1050 n
601 1053 m
592 1086 m
583 1083 n
577 1074 n
574 1059 n
574 1050 n
577 1035 n
583 1026 n
592 1023 n
598 1023 n
607 1026 n
613 1035 n
616 1050 n
616 1059 n
613 1074 n
607 1083 n
598 1086 n
592 1086 n
592 1086 m
586 1083 n
583 1080 n
580 1074 n
577 1059 n
577 1050 n
580 1035 n
583 1029 n
586 1026 n
592 1023 n
598 1023 m
604 1026 n
607 1029 n
610 1035 n
613 1050 n
613 1059 n
610 1074 n
607 1080 n
604 1083 n
598 1086 n
661 1053 m
640 1029 m
637 1026 n
640 1023 n
643 1026 n
640 1029 n
691 1053 m
673 1074 m
679 1077 n
688 1086 n
688 1023 n
685 1083 m
685 1023 n
673 1023 m
700 1023 n
676 1218 m
667 1215 n
661 1206 n
658 1191 n
658 1182 n
661 1167 n
667 1158 n
676 1155 n
682 1155 n
691 1158 n
697 1167 n
700 1182 n
700 1191 n
697 1206 n
691 1215 n
682 1218 n
676 1218 n
676 1218 m
670 1215 n
667 1212 n
664 1206 n
661 1191 n
661 1182 n
664 1167 n
667 1161 n
670 1158 n
676 1155 n
682 1155 m
688 1158 n
691 1161 n
694 1167 n
697 1182 n
697 1191 n
694 1206 n
691 1212 n
688 1215 n
682 1218 n
592 1349 m
583 1346 n
577 1337 n
574 1322 n
574 1313 n
577 1298 n
583 1289 n
592 1286 n
598 1286 n
607 1289 n
613 1298 n
616 1313 n
616 1322 n
613 1337 n
607 1346 n
598 1349 n
592 1349 n
592 1349 m
586 1346 n
583 1343 n
580 1337 n
577 1322 n
577 1313 n
580 1298 n
583 1292 n
586 1289 n
592 1286 n
598 1286 m
604 1289 n
607 1292 n
610 1298 n
613 1313 n
613 1322 n
610 1337 n
607 1343 n
604 1346 n
598 1349 n
661 1316 m
640 1292 m
637 1289 n
640 1286 n
643 1289 n
640 1292 n
691 1316 m
673 1337 m
679 1340 n
688 1349 n
688 1286 n
685 1346 m
685 1286 n
673 1286 m
700 1286 n
586 1481 m
577 1478 n
571 1469 n
568 1454 n
568 1445 n
571 1430 n
577 1421 n
586 1418 n
592 1418 n
601 1421 n
607 1430 n
610 1445 n
610 1454 n
607 1469 n
601 1478 n
592 1481 n
586 1481 n
586 1481 m
580 1478 n
577 1475 n
574 1469 n
571 1454 n
571 1445 n
574 1430 n
577 1424 n
580 1421 n
586 1418 n
592 1418 m
598 1421 n
601 1424 n
604 1430 n
607 1445 n
607 1454 n
604 1469 n
601 1475 n
598 1478 n
592 1481 n
655 1448 m
634 1424 m
631 1421 n
634 1418 n
637 1421 n
634 1424 n
685 1448 m
661 1469 m
664 1466 n
661 1463 n
658 1466 n
658 1469 n
661 1475 n
664 1478 n
673 1481 n
685 1481 n
694 1478 n
697 1475 n
700 1469 n
700 1463 n
697 1457 n
688 1451 n
673 1445 n
667 1442 n
661 1436 n
658 1427 n
658 1418 n
685 1481 m
691 1478 n
694 1475 n
697 1469 n
697 1463 n
694 1457 n
685 1451 n
673 1445 n
658 1424 m
661 1427 n
667 1427 n
682 1421 n
691 1421 n
697 1424 n
700 1427 n
667 1427 m
682 1418 n
694 1418 n
697 1421 n
700 1427 n
700 1433 n
347 1200 m
410 1176 n
347 1200 m
410 1224 n
356 1200 m
410 1221 n
407 1179 m
407 1221 n
410 1176 m
410 1224 n
380 1266 m
347 1245 m
410 1245 n
347 1248 m
410 1248 n
365 1266 m
389 1266 n
347 1236 m
347 1284 n
365 1284 n
347 1281 n
377 1248 m
377 1266 n
410 1236 m
410 1284 n
392 1284 n
410 1281 n
0 0 4096 4096 s
/solid f
/solid f
883 581 p
883 581 p
883 581 p
883 581 p
883 581 p
883 581 p
883 581 p
883 581 p
883 581 p
883 581 p
883 581 p
883 581 p
883 581 p
883 581 p
883 581 p
883 581 p
883 581 p
883 581 p
883 581 p
883 581 p
883 581 p
883 581 p
883 581 p
883 581 p
883 581 p
883 581 p
883 581 p
883 581 p
883 581 p
883 581 p
883 581 p
883 581 p
883 581 p
883 581 p
883 581 p
883 581 p
883 581 p
883 581 p
883 581 p
883 581 p
883 581 p
883 581 p
883 581 p
883 581 p
883 581 p
883 581 p
883 581 p
883 581 p
883 581 p
883 581 p
883 581 p
883 581 p
883 581 p
883 581 p
883 581 p
883 581 p
883 581 p
883 581 p
883 581 p
883 581 p
883 581 p
883 581 p
883 581 p
883 581 p
883 581 p
883 581 p
883 581 p
883 581 p
883 581 p
883 581 p
883 581 p
883 581 p
883 581 p
883 581 p
883 581 p
883 581 p
883 581 p
883 581 p
883 581 p
883 581 p
883 581 p
883 581 p
883 581 p
883 581 p
883 581 p
883 581 p
883 581 p
883 581 p
883 581 p
883 581 p
883 581 p
883 581 p
883 581 p
883 581 p
883 581 p
883 581 p
883 581 p
883 581 p
883 581 p
883 581 p
883 581 p
883 581 p
883 581 p
883 581 p
883 581 p
883 581 p
883 581 p
883 581 p
883 581 p
883 581 p
883 581 p
883 581 p
883 581 p
883 581 p
883 581 p
883 581 p
883 581 p
883 581 p
883 581 p
883 581 p
883 581 p
883 581 p
883 581 p
883 581 p
883 581 p
883 581 p
883 581 p
883 581 p
883 581 p
883 581 p
883 581 p
883 581 p
883 581 p
883 581 p
883 581 p
883 581 p
883 581 p
883 581 p
883 581 p
883 581 p
883 581 p
883 581 p
883 581 p
883 581 p
883 581 p
883 581 p
883 581 p
883 581 p
883 581 p
883 581 p
883 581 p
883 581 p
883 581 p
883 581 p
883 581 p
883 581 p
883 581 p
883 581 p
883 581 p
883 581 p
883 581 p
883 581 p
883 581 p
883 581 p
883 581 p
883 581 p
883 581 p
883 581 p
883 581 p
883 581 p
883 581 p
883 581 p
883 581 p
883 581 p
883 581 p
883 581 p
883 581 p
883 581 p
883 581 p
883 581 p
883 581 p
883 581 p
883 581 p
883 581 p
883 581 p
883 581 p
883 581 p
883 581 p
883 581 p
883 581 p
883 581 p
883 581 p
883 581 p
883 581 p
883 581 p
883 581 p
883 581 p
883 581 p
883 581 p
883 581 p
883 581 p
883 581 p
883 581 p
883 581 p
883 581 p
883 581 p
883 581 p
883 581 p
883 581 p
883 581 p
883 581 p
883 581 p
883 581 p
883 581 p
883 581 p
883 581 p
883 581 p
883 581 p
883 581 p
883 581 p
883 581 p
883 581 p
883 581 p
883 581 p
883 581 p
883 581 p
883 581 p
883 581 p
883 581 p
883 581 p
883 581 p
883 581 p
883 581 p
883 581 p
883 581 p
883 581 p
883 581 p
883 581 p
883 581 p
883 581 p
883 581 p
883 581 p
883 581 p
883 581 p
883 581 p
883 581 p
883 581 p
883 581 p
883 581 p
883 581 p
883 581 p
883 581 p
883 581 p
883 581 p
883 581 p
883 581 p
883 581 p
883 581 p
883 581 p
883 581 p
883 581 p
883 581 p
883 581 p
883 581 p
883 581 p
883 581 p
883 581 p
883 581 p
883 581 p
883 581 p
883 581 p
883 581 p
883 581 p
883 581 p
883 581 p
883 581 p
883 581 p
883 581 p
883 581 p
883 581 p
883 581 p
883 581 p
883 581 p
883 581 p
883 581 p
883 581 p
883 581 p
883 581 p
883 581 p
883 581 p
883 581 p
883 581 p
883 581 p
883 581 p
883 581 p
883 581 p
883 581 p
883 581 p
883 581 p
883 581 p
883 581 p
883 581 p
883 581 p
883 581 p
883 581 p
883 581 p
883 581 p
883 581 p
883 581 p
883 581 p
883 581 p
883 581 p
883 581 p
883 581 p
883 581 p
883 581 p
883 581 p
883 581 p
883 581 p
883 581 p
883 581 p
883 581 p
883 581 p
883 581 p
883 581 p
883 581 p
883 581 p
883 581 p
883 581 p
883 581 p
883 581 p
883 581 p
/solid f
896 744 m
896 744 m
/shortdashed f
883 581 m
1016 698 n
1149 789 n
1282 864 n
1415 918 n
1548 938 n
1681 933 n
1814 920 n
1947 908 n
2080 898 n
2212 892 n
2345 888 n
2478 886 n
2611 885 n
2744 885 n
2877 886 n
3010 887 n
3143 889 n
3276 891 n
3409 893 n
3542 896 n
/solid f
883 581 m
1016 565 n
1149 549 n
1282 565 n
1415 613 n
1548 665 n
1681 705 n
1814 735 n
1947 760 n
2080 782 n
2212 802 n
2345 820 n
2478 838 n
2611 854 n
2744 870 n
2877 885 n
3010 898 n
3143 911 n
3276 924 n
3409 935 n
3542 947 n
/solid f
750 500 m
3675 500 n
3675 987 n
750 987 n
750 500 n
883 500 m
883 600 n
883 887 m
883 987 n
1149 500 m
1149 550 n
1149 937 m
1149 987 n
1415 500 m
1415 550 n
1415 937 m
1415 987 n
1681 500 m
1681 550 n
1681 937 m
1681 987 n
1947 500 m
1947 550 n
1947 937 m
1947 987 n
2212 500 m
2212 600 n
2212 887 m
2212 987 n
2478 500 m
2478 550 n
2478 937 m
2478 987 n
2744 500 m
2744 550 n
2744 937 m
2744 987 n
3010 500 m
3010 550 n
3010 937 m
3010 987 n
3276 500 m
3276 550 n
3276 937 m
3276 987 n
3542 500 m
3542 600 n
3542 887 m
3542 987 n
750 516 m
800 516 n
3625 516 m
3675 516 n
750 549 m
800 549 n
3625 549 m
3675 549 n
750 581 m
850 581 n
3575 581 m
3675 581 n
750 614 m
800 614 n
3625 614 m
3675 614 n
750 646 m
800 646 n
3625 646 m
3675 646 n
750 679 m
800 679 n
3625 679 m
3675 679 n
750 711 m
800 711 n
3625 711 m
3675 711 n
750 744 m
850 744 n
3575 744 m
3675 744 n
750 776 m
800 776 n
3625 776 m
3675 776 n
750 809 m
800 809 n
3625 809 m
3675 809 n
750 841 m
800 841 n
3625 841 m
3675 841 n
750 874 m
800 874 n
3625 874 m
3675 874 n
750 906 m
850 906 n
3575 906 m
3675 906 n
750 939 m
800 939 n
3625 939 m
3675 939 n
750 971 m
800 971 n
3625 971 m
3675 971 n
943 958 m
937 952 n
931 943 n
925 931 n
922 916 n
922 904 n
925 889 n
931 877 n
937 868 n
943 862 n
937 952 m
931 940 n
928 931 n
925 916 n
925 904 n
928 889 n
931 880 n
937 868 n
988 913 m
976 925 m
970 922 n
967 919 n
964 913 n
964 907 n
967 901 n
970 898 n
976 895 n
982 895 n
988 898 n
991 901 n
994 907 n
994 913 n
991 919 n
988 922 n
982 925 n
976 925 n
970 922 m
967 916 n
967 904 n
970 898 n
988 898 m
991 904 n
991 916 n
988 922 n
991 919 m
994 922 n
1000 925 n
1000 922 n
994 922 n
967 901 m
964 898 n
961 892 n
961 889 n
964 883 n
973 880 n
988 880 n
997 877 n
1000 874 n
961 889 m
964 886 n
973 883 n
988 883 n
997 880 n
1000 874 n
1000 871 n
997 865 n
988 862 n
970 862 n
961 865 n
958 871 n
958 874 n
961 880 n
970 883 n
1045 913 m
1018 958 m
1024 952 n
1030 943 n
1036 931 n
1039 916 n
1039 904 n
1036 889 n
1030 877 n
1024 868 n
1018 862 n
1024 952 m
1030 940 n
1033 931 n
1036 916 n
1036 904 n
1033 889 n
1030 880 n
1024 868 n
880 413 m
871 410 n
865 401 n
862 386 n
862 377 n
865 362 n
871 353 n
880 350 n
886 350 n
895 353 n
901 362 n
904 377 n
904 386 n
901 401 n
895 410 n
886 413 n
880 413 n
880 413 m
874 410 n
871 407 n
868 401 n
865 386 n
865 377 n
868 362 n
871 356 n
874 353 n
880 350 n
886 350 m
892 353 n
895 356 n
898 362 n
901 377 n
901 386 n
898 401 n
895 407 n
892 410 n
886 413 n
2165 401 m
2171 404 n
2180 413 n
2180 350 n
2177 410 m
2177 350 n
2165 350 m
2192 350 n
2243 380 m
2234 413 m
2225 410 n
2219 401 n
2216 386 n
2216 377 n
2219 362 n
2225 353 n
2234 350 n
2240 350 n
2249 353 n
2255 362 n
2258 377 n
2258 386 n
2255 401 n
2249 410 n
2240 413 n
2234 413 n
2234 413 m
2228 410 n
2225 407 n
2222 401 n
2219 386 n
2219 377 n
2222 362 n
2225 356 n
2228 353 n
2234 350 n
2240 350 m
2246 353 n
2249 356 n
2252 362 n
2255 377 n
2255 386 n
2252 401 n
2249 407 n
2246 410 n
2240 413 n
3494 401 m
3497 398 n
3494 395 n
3491 398 n
3491 401 n
3494 407 n
3497 410 n
3506 413 n
3518 413 n
3527 410 n
3530 407 n
3533 401 n
3533 395 n
3530 389 n
3521 383 n
3506 377 n
3500 374 n
3494 368 n
3491 359 n
3491 350 n
3518 413 m
3524 410 n
3527 407 n
3530 401 n
3530 395 n
3527 389 n
3518 383 n
3506 377 n
3491 356 m
3494 359 n
3500 359 n
3515 353 n
3524 353 n
3530 356 n
3533 359 n
3500 359 m
3515 350 n
3527 350 n
3530 353 n
3533 359 n
3533 365 n
3578 380 m
3569 413 m
3560 410 n
3554 401 n
3551 386 n
3551 377 n
3554 362 n
3560 353 n
3569 350 n
3575 350 n
3584 353 n
3590 362 n
3593 377 n
3593 386 n
3590 401 n
3584 410 n
3575 413 n
3569 413 n
3569 413 m
3563 410 n
3560 407 n
3557 401 n
3554 386 n
3554 377 n
3557 362 n
3560 356 n
3563 353 n
3569 350 n
3575 350 m
3581 353 n
3584 356 n
3587 362 n
3590 377 n
3590 386 n
3587 401 n
3584 407 n
3581 410 n
3575 413 n
676 614 m
667 611 n
661 602 n
658 587 n
658 578 n
661 563 n
667 554 n
676 551 n
682 551 n
691 554 n
697 563 n
700 578 n
700 587 n
697 602 n
691 611 n
682 614 n
676 614 n
676 614 m
670 611 n
667 608 n
664 602 n
661 587 n
661 578 n
664 563 n
667 557 n
670 554 n
676 551 n
682 551 m
688 554 n
691 557 n
694 563 n
697 578 n
697 587 n
694 602 n
691 608 n
688 611 n
682 614 n
592 777 m
583 774 n
577 765 n
574 750 n
574 741 n
577 726 n
583 717 n
592 714 n
598 714 n
607 717 n
613 726 n
616 741 n
616 750 n
613 765 n
607 774 n
598 777 n
592 777 n
592 777 m
586 774 n
583 771 n
580 765 n
577 750 n
577 741 n
580 726 n
583 720 n
586 717 n
592 714 n
598 714 m
604 717 n
607 720 n
610 726 n
613 741 n
613 750 n
610 765 n
607 771 n
604 774 n
598 777 n
661 744 m
640 720 m
637 717 n
640 714 n
643 717 n
640 720 n
691 744 m
673 765 m
679 768 n
688 777 n
688 714 n
685 774 m
685 714 n
673 714 m
700 714 n
586 939 m
577 936 n
571 927 n
568 912 n
568 903 n
571 888 n
577 879 n
586 876 n
592 876 n
601 879 n
607 888 n
610 903 n
610 912 n
607 927 n
601 936 n
592 939 n
586 939 n
586 939 m
580 936 n
577 933 n
574 927 n
571 912 n
571 903 n
574 888 n
577 882 n
580 879 n
586 876 n
592 876 m
598 879 n
601 882 n
604 888 n
607 903 n
607 912 n
604 927 n
601 933 n
598 936 n
592 939 n
655 906 m
634 882 m
631 879 n
634 876 n
637 879 n
634 882 n
685 906 m
661 927 m
664 924 n
661 921 n
658 924 n
658 927 n
661 933 n
664 936 n
673 939 n
685 939 n
694 936 n
697 933 n
700 927 n
700 921 n
697 915 n
688 909 n
673 903 n
667 900 n
661 894 n
658 885 n
658 876 n
685 939 m
691 936 n
694 933 n
697 927 n
697 921 n
694 915 n
685 909 n
673 903 n
658 882 m
661 885 n
667 885 n
682 879 n
691 879 n
697 882 n
700 885 n
667 885 m
682 876 n
694 876 n
697 879 n
700 885 n
700 891 n
2630 293 m
2630 248 n
2633 239 n
2639 233 n
2648 230 n
2654 230 n
2663 233 n
2669 239 n
2672 248 n
2672 293 n
2633 293 m
2633 248 n
2636 239 n
2642 233 n
2648 230 n
2621 293 m
2642 293 n
2663 293 m
2681 293 n
407 713 m
470 689 n
407 713 m
470 737 n
416 713 m
470 734 n
467 692 m
467 734 n
470 689 m
470 737 n
440 779 m
407 758 m
470 758 n
407 761 m
470 761 n
425 779 m
449 779 n
407 749 m
407 797 n
425 797 n
407 794 n
437 761 m
437 779 n
470 749 m
470 797 n
452 797 n
470 794 n
0 0 4096 4096 s
/solid f
/solid f
883 581 p
883 581 p
883 581 p
883 581 p
883 581 p
883 581 p
883 581 p
883 581 p
883 581 p
883 581 p
883 581 p
883 581 p
883 581 p
883 581 p
883 581 p
883 581 p
883 581 p
883 581 p
883 581 p
883 581 p
883 581 p
883 581 p
883 581 p
883 581 p
883 581 p
883 581 p
883 581 p
883 581 p
883 581 p
883 581 p
883 581 p
883 581 p
883 581 p
883 581 p
883 581 p
883 581 p
883 581 p
883 581 p
883 581 p
883 581 p
883 581 p
883 581 p
883 581 p
883 581 p
883 581 p
883 581 p
883 581 p
883 581 p
883 581 p
883 581 p
883 581 p
883 581 p
883 581 p
883 581 p
883 581 p
883 581 p
883 581 p
883 581 p
883 581 p
883 581 p
883 581 p
883 581 p
883 581 p
883 581 p
883 581 p
883 581 p
883 581 p
883 581 p
883 581 p
883 581 p
883 581 p
883 581 p
883 581 p
883 581 p
883 581 p
883 581 p
883 581 p
883 581 p
883 581 p
883 581 p
883 581 p
883 581 p
883 581 p
883 581 p
883 581 p
883 581 p
883 581 p
883 581 p
883 581 p
883 581 p
883 581 p
883 581 p
883 581 p
883 581 p
883 581 p
883 581 p
883 581 p
883 581 p
883 581 p
883 581 p
883 581 p
883 581 p
883 581 p
883 581 p
883 581 p
883 581 p
883 581 p
883 581 p
883 581 p
883 581 p
883 581 p
883 581 p
883 581 p
883 581 p
883 581 p
883 581 p
883 581 p
883 581 p
883 581 p
883 581 p
883 581 p
883 581 p
883 581 p
883 581 p
883 581 p
883 581 p
883 581 p
883 581 p
883 581 p
883 581 p
883 581 p
883 581 p
883 581 p
883 581 p
883 581 p
883 581 p
883 581 p
883 581 p
883 581 p
883 581 p
883 581 p
883 581 p
883 581 p
883 581 p
883 581 p
883 581 p
883 581 p
883 581 p
883 581 p
883 581 p
883 581 p
883 581 p
883 581 p
883 581 p
883 581 p
883 581 p
883 581 p
883 581 p
883 581 p
883 581 p
883 581 p
883 581 p
883 581 p
883 581 p
883 581 p
883 581 p
883 581 p
883 581 p
883 581 p
883 581 p
883 581 p
883 581 p
883 581 p
883 581 p
883 581 p
883 581 p
883 581 p
883 581 p
883 581 p
883 581 p
883 581 p
883 581 p
883 581 p
883 581 p
883 581 p
883 581 p
883 581 p
883 581 p
883 581 p
883 581 p
883 581 p
883 581 p
883 581 p
883 581 p
883 581 p
883 581 p
883 581 p
883 581 p
883 581 p
883 581 p
883 581 p
883 581 p
883 581 p
883 581 p
883 581 p
883 581 p
883 581 p
883 581 p
883 581 p
883 581 p
883 581 p
883 581 p
883 581 p
883 581 p
883 581 p
883 581 p
883 581 p
883 581 p
883 581 p
883 581 p
883 581 p
883 581 p
883 581 p
883 581 p
883 581 p
883 581 p
883 581 p
883 581 p
883 581 p
883 581 p
883 581 p
883 581 p
883 581 p
883 581 p
883 581 p
883 581 p
883 581 p
883 581 p
883 581 p
883 581 p
883 581 p
883 581 p
883 581 p
883 581 p
883 581 p
883 581 p
883 581 p
883 581 p
883 581 p
883 581 p
883 581 p
883 581 p
883 581 p
883 581 p
883 581 p
883 581 p
883 581 p
883 581 p
883 581 p
883 581 p
883 581 p
883 581 p
883 581 p
883 581 p
883 581 p
883 581 p
883 581 p
883 581 p
883 581 p
883 581 p
883 581 p
883 581 p
883 581 p
883 581 p
883 581 p
883 581 p
883 581 p
883 581 p
883 581 p
883 581 p
883 581 p
883 581 p
883 581 p
883 581 p
883 581 p
883 581 p
883 581 p
883 581 p
883 581 p
883 581 p
883 581 p
883 581 p
883 581 p
883 581 p
883 581 p
883 581 p
883 581 p
883 581 p
883 581 p
883 581 p
883 581 p
883 581 p
883 581 p
883 581 p
883 581 p
883 581 p
883 581 p
883 581 p
883 581 p
883 581 p
883 581 p
883 581 p
883 581 p
883 581 p
883 581 p
883 581 p
883 581 p
883 581 p
883 581 p
883 581 p
883 581 p
883 581 p
883 581 p
883 581 p
883 581 p
883 581 p
883 581 p
883 581 p
883 581 p
883 581 p
883 581 p
883 581 p
883 581 p
883 581 p
883 581 p
883 581 p
883 581 p
883 581 p
883 581 p
883 581 p
883 581 p
883 581 p
883 581 p
883 581 p
883 581 p
883 581 p
883 581 p
883 581 p
883 581 p
883 581 p
883 581 p
883 581 p
883 581 p
883 581 p
883 581 p
883 581 p
883 581 p
883 581 p
883 581 p
883 581 p
883 581 p
883 581 p
883 581 p
883 581 p
883 581 p
883 581 p
883 581 p
883 581 p
883 581 p
883 581 p
883 581 p
883 581 p
883 581 p
883 581 p
883 581 p
/solid f
896 744 m
896 744 m
/solid f
750 500 m
3675 500 n
3675 987 n
750 987 n
750 500 n
883 500 m
883 600 n
883 887 m
883 987 n
1149 500 m
1149 550 n
1149 937 m
1149 987 n
1415 500 m
1415 550 n
1415 937 m
1415 987 n
1681 500 m
1681 550 n
1681 937 m
1681 987 n
1947 500 m
1947 550 n
1947 937 m
1947 987 n
2212 500 m
2212 600 n
2212 887 m
2212 987 n
2478 500 m
2478 550 n
2478 937 m
2478 987 n
2744 500 m
2744 550 n
2744 937 m
2744 987 n
3010 500 m
3010 550 n
3010 937 m
3010 987 n
3276 500 m
3276 550 n
3276 937 m
3276 987 n
3542 500 m
3542 600 n
3542 887 m
3542 987 n
750 516 m
800 516 n
3625 516 m
3675 516 n
750 549 m
800 549 n
3625 549 m
3675 549 n
750 581 m
850 581 n
3575 581 m
3675 581 n
750 614 m
800 614 n
3625 614 m
3675 614 n
750 646 m
800 646 n
3625 646 m
3675 646 n
750 679 m
800 679 n
3625 679 m
3675 679 n
750 711 m
800 711 n
3625 711 m
3675 711 n
750 744 m
850 744 n
3575 744 m
3675 744 n
750 776 m
800 776 n
3625 776 m
3675 776 n
750 809 m
800 809 n
3625 809 m
3675 809 n
750 841 m
800 841 n
3625 841 m
3675 841 n
750 874 m
800 874 n
3625 874 m
3675 874 n
750 906 m
850 906 n
3575 906 m
3675 906 n
750 939 m
800 939 n
3625 939 m
3675 939 n
750 971 m
800 971 n
3625 971 m
3675 971 n
1676 80 m
1678 73 n
1678 87 n
1676 80 n
1671 85 n
1664 87 n
1660 87 n
1652 85 n
1648 80 n
1645 75 n
1643 68 n
1643 57 n
1645 49 n
1648 45 n
1652 40 n
1660 37 n
1664 37 n
1671 40 n
1676 45 n
1660 87 m
1655 85 n
1650 80 n
1648 75 n
1645 68 n
1645 57 n
1648 49 n
1650 45 n
1655 40 n
1660 37 n
1676 57 m
1676 37 n
1678 57 m
1678 37 n
1668 57 m
1685 57 n
1719 61 m
1702 65 m
1702 63 n
1700 63 n
1700 65 n
1702 68 n
1707 70 n
1716 70 n
1721 68 n
1724 65 n
1726 61 n
1726 45 n
1728 40 n
1731 37 n
1724 65 m
1724 45 n
1726 40 n
1731 37 n
1733 37 n
1724 61 m
1721 59 n
1707 57 n
1700 54 n
1697 49 n
1697 45 n
1700 40 n
1707 37 n
1714 37 n
1719 40 n
1724 45 n
1707 57 m
1702 54 n
1700 49 n
1700 45 n
1702 40 n
1707 37 n
1767 61 m
1750 70 m
1750 37 n
1752 70 m
1752 37 n
1752 63 m
1757 68 n
1764 70 n
1769 70 n
1776 68 n
1779 63 n
1779 37 n
1769 70 m
1774 68 n
1776 63 n
1776 37 n
1779 63 m
1784 68 n
1791 70 n
1796 70 n
1803 68 n
1805 63 n
1805 37 n
1796 70 m
1800 68 n
1803 63 n
1803 37 n
1743 70 m
1752 70 n
1743 37 m
1760 37 n
1769 37 m
1786 37 n
1796 37 m
1812 37 n
1846 61 m
1829 70 m
1829 37 n
1832 70 m
1832 37 n
1832 63 m
1836 68 n
1844 70 n
1848 70 n
1856 68 n
1858 63 n
1858 37 n
1848 70 m
1853 68 n
1856 63 n
1856 37 n
1858 63 m
1863 68 n
1870 70 n
1875 70 n
1882 68 n
1884 63 n
1884 37 n
1875 70 m
1880 68 n
1882 63 n
1882 37 n
1822 70 m
1832 70 n
1822 37 m
1839 37 n
1848 37 m
1865 37 n
1875 37 m
1892 37 n
1925 61 m
1906 57 m
1935 57 n
1935 61 n
1932 65 n
1930 68 n
1925 70 n
1918 70 n
1911 68 n
1906 63 n
1904 57 n
1904 52 n
1906 45 n
1911 40 n
1918 37 n
1923 37 n
1930 40 n
1935 45 n
1932 57 m
1932 63 n
1930 68 n
1918 70 m
1913 68 n
1908 63 n
1906 57 n
1906 52 n
1908 45 n
1913 40 n
1918 37 n
1971 61 m
1954 87 m
1954 37 n
1956 87 m
1956 37 n
1947 87 m
1956 87 n
1947 37 m
1964 37 n
1997 61 m
2036 61 m
2016 57 m
2045 57 n
2045 61 n
2043 65 n
2040 68 n
2036 70 n
2028 70 n
2021 68 n
2016 63 n
2014 57 n
2014 52 n
2016 45 n
2021 40 n
2028 37 n
2033 37 n
2040 40 n
2045 45 n
2043 57 m
2043 63 n
2040 68 n
2028 70 m
2024 68 n
2019 63 n
2016 57 n
2016 52 n
2019 45 n
2024 40 n
2028 37 n
2081 61 m
2064 87 m
2064 47 n
2067 40 n
2072 37 n
2076 37 n
2081 40 n
2084 45 n
2067 87 m
2067 47 n
2069 40 n
2072 37 n
2057 70 m
2076 70 n
2117 61 m
2156 61 m
2139 65 m
2139 63 n
2136 63 n
2136 65 n
2139 68 n
2144 70 n
2153 70 n
2158 68 n
2160 65 n
2163 61 n
2163 45 n
2165 40 n
2168 37 n
2160 65 m
2160 45 n
2163 40 n
2168 37 n
2170 37 n
2160 61 m
2158 59 n
2144 57 n
2136 54 n
2134 49 n
2134 45 n
2136 40 n
2144 37 n
2151 37 n
2156 40 n
2160 45 n
2144 57 m
2139 54 n
2136 49 n
2136 45 n
2139 40 n
2144 37 n
2204 61 m
2187 87 m
2187 37 n
2189 87 m
2189 37 n
2180 87 m
2189 87 n
2180 37 m
2196 37 n
2230 61 m
2268 61 m
2252 87 m
2252 37 n
2254 87 m
2254 37 n
2268 73 m
2268 54 n
2244 87 m
2283 87 n
2283 73 n
2280 87 n
2254 63 m
2268 63 n
2244 37 m
2261 37 n
2316 61 m
2300 87 m
2297 85 n
2300 82 n
2302 85 n
2300 87 n
2300 70 m
2300 37 n
2302 70 m
2302 37 n
2292 70 m
2302 70 n
2292 37 m
2309 37 n
2343 61 m
2333 70 m
2328 68 n
2326 65 n
2324 61 n
2324 57 n
2326 52 n
2328 49 n
2333 47 n
2338 47 n
2343 49 n
2345 52 n
2348 57 n
2348 61 n
2345 65 n
2343 68 n
2338 70 n
2333 70 n
2328 68 m
2326 63 n
2326 54 n
2328 49 n
2343 49 m
2345 54 n
2345 63 n
2343 68 n
2345 65 m
2348 68 n
2352 70 n
2352 68 n
2348 68 n
2326 52 m
2324 49 n
2321 45 n
2321 42 n
2324 37 n
2331 35 n
2343 35 n
2350 33 n
2352 30 n
2321 42 m
2324 40 n
2331 37 n
2343 37 n
2350 35 n
2352 30 n
2352 28 n
2350 23 n
2343 21 n
2328 21 n
2321 23 n
2319 28 n
2319 30 n
2321 35 n
2328 37 n
2388 61 m
2372 42 m
2369 40 n
2372 37 n
2374 40 n
2372 42 n
2412 61 m
2415 65 m
2417 70 n
2417 61 n
2415 65 n
2412 68 n
2408 70 n
2398 70 n
2393 68 n
2391 65 n
2391 61 n
2393 59 n
2398 57 n
2410 52 n
2415 49 n
2417 47 n
2391 63 m
2393 61 n
2398 59 n
2410 54 n
2415 52 n
2417 49 n
2417 42 n
2415 40 n
2410 37 n
2400 37 n
2396 40 n
2393 42 n
2391 47 n
2391 37 n
2393 42 n
2453 61 m
2492 61 m
2477 77 m
2482 80 n
2489 87 n
2489 37 n
2487 85 m
2487 37 n
2477 37 m
2499 37 n
2540 61 m
2578 61 m
2561 65 m
2561 63 n
2559 63 n
2559 65 n
2561 68 n
2566 70 n
2576 70 n
2580 68 n
2583 65 n
2585 61 n
2585 45 n
2588 40 n
2590 37 n
2583 65 m
2583 45 n
2585 40 n
2590 37 n
2592 37 n
2583 61 m
2580 59 n
2566 57 n
2559 54 n
2556 49 n
2556 45 n
2559 40 n
2566 37 n
2573 37 n
2578 40 n
2583 45 n
2566 57 m
2561 54 n
2559 49 n
2559 45 n
2561 40 n
2566 37 n
2626 61 m
2609 70 m
2609 37 n
2612 70 m
2612 37 n
2612 63 m
2616 68 n
2624 70 n
2628 70 n
2636 68 n
2638 63 n
2638 37 n
2628 70 m
2633 68 n
2636 63 n
2636 37 n
2602 70 m
2612 70 n
2602 37 m
2619 37 n
2628 37 m
2645 37 n
2679 61 m
2686 87 m
2686 37 n
2688 87 m
2688 37 n
2686 63 m
2681 68 n
2676 70 n
2672 70 n
2664 68 n
2660 63 n
2657 57 n
2657 52 n
2660 45 n
2664 40 n
2672 37 n
2676 37 n
2681 40 n
2686 45 n
2672 70 m
2667 68 n
2662 63 n
2660 57 n
2660 52 n
2662 45 n
2667 40 n
2672 37 n
2679 87 m
2688 87 n
2686 37 m
2696 37 n
2729 61 m
2768 61 m
2748 77 m
2751 75 n
2748 73 n
2746 75 n
2746 77 n
2748 82 n
2751 85 n
2758 87 n
2768 87 n
2775 85 n
2777 82 n
2780 77 n
2780 73 n
2777 68 n
2770 63 n
2758 59 n
2753 57 n
2748 52 n
2746 45 n
2746 37 n
2768 87 m
2772 85 n
2775 82 n
2777 77 n
2777 73 n
2775 68 n
2768 63 n
2758 59 n
2746 42 m
2748 45 n
2753 45 n
2765 40 n
2772 40 n
2777 42 n
2780 45 n
2753 45 m
2765 37 n
2775 37 n
2777 40 n
2780 45 n
2780 49 n
880 413 m
871 410 n
865 401 n
862 386 n
862 377 n
865 362 n
871 353 n
880 350 n
886 350 n
895 353 n
901 362 n
904 377 n
904 386 n
901 401 n
895 410 n
886 413 n
880 413 n
880 413 m
874 410 n
871 407 n
868 401 n
865 386 n
865 377 n
868 362 n
871 356 n
874 353 n
880 350 n
886 350 m
892 353 n
895 356 n
898 362 n
901 377 n
901 386 n
898 401 n
895 407 n
892 410 n
886 413 n
2165 401 m
2171 404 n
2180 413 n
2180 350 n
2177 410 m
2177 350 n
2165 350 m
2192 350 n
2243 380 m
2234 413 m
2225 410 n
2219 401 n
2216 386 n
2216 377 n
2219 362 n
2225 353 n
2234 350 n
2240 350 n
2249 353 n
2255 362 n
2258 377 n
2258 386 n
2255 401 n
2249 410 n
2240 413 n
2234 413 n
2234 413 m
2228 410 n
2225 407 n
2222 401 n
2219 386 n
2219 377 n
2222 362 n
2225 356 n
2228 353 n
2234 350 n
2240 350 m
2246 353 n
2249 356 n
2252 362 n
2255 377 n
2255 386 n
2252 401 n
2249 407 n
2246 410 n
2240 413 n
3494 401 m
3497 398 n
3494 395 n
3491 398 n
3491 401 n
3494 407 n
3497 410 n
3506 413 n
3518 413 n
3527 410 n
3530 407 n
3533 401 n
3533 395 n
3530 389 n
3521 383 n
3506 377 n
3500 374 n
3494 368 n
3491 359 n
3491 350 n
3518 413 m
3524 410 n
3527 407 n
3530 401 n
3530 395 n
3527 389 n
3518 383 n
3506 377 n
3491 356 m
3494 359 n
3500 359 n
3515 353 n
3524 353 n
3530 356 n
3533 359 n
3500 359 m
3515 350 n
3527 350 n
3530 353 n
3533 359 n
3533 365 n
3578 380 m
3569 413 m
3560 410 n
3554 401 n
3551 386 n
3551 377 n
3554 362 n
3560 353 n
3569 350 n
3575 350 n
3584 353 n
3590 362 n
3593 377 n
3593 386 n
3590 401 n
3584 410 n
3575 413 n
3569 413 n
3569 413 m
3563 410 n
3560 407 n
3557 401 n
3554 386 n
3554 377 n
3557 362 n
3560 356 n
3563 353 n
3569 350 n
3575 350 m
3581 353 n
3584 356 n
3587 362 n
3590 377 n
3590 386 n
3587 401 n
3584 407 n
3581 410 n
3575 413 n
676 614 m
667 611 n
661 602 n
658 587 n
658 578 n
661 563 n
667 554 n
676 551 n
682 551 n
691 554 n
697 563 n
700 578 n
700 587 n
697 602 n
691 611 n
682 614 n
676 614 n
676 614 m
670 611 n
667 608 n
664 602 n
661 587 n
661 578 n
664 563 n
667 557 n
670 554 n
676 551 n
682 551 m
688 554 n
691 557 n
694 563 n
697 578 n
697 587 n
694 602 n
691 608 n
688 611 n
682 614 n
592 777 m
583 774 n
577 765 n
574 750 n
574 741 n
577 726 n
583 717 n
592 714 n
598 714 n
607 717 n
613 726 n
616 741 n
616 750 n
613 765 n
607 774 n
598 777 n
592 777 n
592 777 m
586 774 n
583 771 n
580 765 n
577 750 n
577 741 n
580 726 n
583 720 n
586 717 n
592 714 n
598 714 m
604 717 n
607 720 n
610 726 n
613 741 n
613 750 n
610 765 n
607 771 n
604 774 n
598 777 n
661 744 m
640 720 m
637 717 n
640 714 n
643 717 n
640 720 n
691 744 m
673 765 m
679 768 n
688 777 n
688 714 n
685 774 m
685 714 n
673 714 m
700 714 n
586 939 m
577 936 n
571 927 n
568 912 n
568 903 n
571 888 n
577 879 n
586 876 n
592 876 n
601 879 n
607 888 n
610 903 n
610 912 n
607 927 n
601 936 n
592 939 n
586 939 n
586 939 m
580 936 n
577 933 n
574 927 n
571 912 n
571 903 n
574 888 n
577 882 n
580 879 n
586 876 n
592 876 m
598 879 n
601 882 n
604 888 n
607 903 n
607 912 n
604 927 n
601 933 n
598 936 n
592 939 n
655 906 m
634 882 m
631 879 n
634 876 n
637 879 n
634 882 n
685 906 m
661 927 m
664 924 n
661 921 n
658 924 n
658 927 n
661 933 n
664 936 n
673 939 n
685 939 n
694 936 n
697 933 n
700 927 n
700 921 n
697 915 n
688 909 n
673 903 n
667 900 n
661 894 n
658 885 n
658 876 n
685 939 m
691 936 n
694 933 n
697 927 n
697 921 n
694 915 n
685 909 n
673 903 n
658 882 m
661 885 n
667 885 n
682 879 n
691 879 n
697 882 n
700 885 n
667 885 m
682 876 n
694 876 n
697 879 n
700 885 n
700 891 n
2630 293 m
2630 248 n
2633 239 n
2639 233 n
2648 230 n
2654 230 n
2663 233 n
2669 239 n
2672 248 n
2672 293 n
2633 293 m
2633 248 n
2636 239 n
2642 233 n
2648 230 n
2621 293 m
2642 293 n
2663 293 m
2681 293 n
407 713 m
470 689 n
407 713 m
470 737 n
416 713 m
470 734 n
467 692 m
467 734 n
470 689 m
470 737 n
440 779 m
407 758 m
470 758 n
407 761 m
470 761 n
425 779 m
449 779 n
407 749 m
407 797 n
425 797 n
407 794 n
437 761 m
437 779 n
470 749 m
470 797 n
452 797 n
470 794 n
e
EndPSPlot
save 50 dict begin /psplot exch def
/StartPSPlot
   {newpath 0 0 moveto 6 setlinewidth 0 setgray 1 setlinecap
    /imtx matrix currentmatrix def /dmtx matrix defaultmatrix def
    /fnt /Courier findfont def /smtx matrix def fnt 8 scalefont setfont}def
/solid {{}0}def
/dotted	{[2 nail 10 nail ] 0}def
/longdashed {[10 nail] 0}def
/shortdashed {[6 nail] 0}def
/dotdashed {[2 nail 6 nail 10 nail 6 nail] 0}def
/min {2 copy lt{pop}{exch pop}ifelse}def
/max {2 copy lt{exch pop}{pop}ifelse}def
/len {dup mul exch dup mul add sqrt}def
/nail {0 imtx dtransform len 0 idtransform len}def

/m {newpath moveto}def
/n {lineto currentpoint stroke moveto}def
/p {newpath moveto gsave 1 setlinecap solid setdash
    dmtx setmatrix .4 nail setlinewidth
    .05 0 idtransform rlineto stroke grestore}def
/l {moveto lineto currentpoint stroke moveto}def
/t {smtx currentmatrix pop imtx setmatrix show smtx setmatrix}def
/a {gsave newpath /y2 exch def /x2 exch def
    /y1 exch def /x1 exch def /yc exch def /xc exch def
    /r x1 xc sub dup mul y1 yc sub dup mul add sqrt
       x2 xc sub dup mul y2 yc sub dup mul add sqrt add 2 div def
    /ang1 y1 yc sub x1 xc sub atan def
    /ang2 y2 yc sub x2 xc sub atan def
    xc yc r ang1 ang2 arc stroke grestore}def
/c {gsave newpath 0 360 arc stroke grestore}def
/e {gsave showpage grestore newpath 0 0 moveto}def
/f {load exec setdash}def
/s {/ury exch def /urx exch def /lly exch def /llx exch def
    imtx setmatrix newpath clippath pathbbox newpath
    /dury exch def /durx exch def /dlly exch def /dllx exch def
    /md durx dllx sub dury dlly sub min def
    /Mu urx llx sub ury lly sub max def
    dllx dlly translate md Mu div dup scale llx neg lly neg translate}def
/EndPSPlot {clear psplot end restore}def

StartPSPlot
0 0 4096 4096 s
/solid f
/solid f
0 0 4096 4096 s
/solid f
/solid f
587 3790 m
587 3790 m
912 3758 m
888 3752 m
924 3752 n
924 3758 n
921 3764 n
918 3767 n
912 3770 n
903 3770 n
894 3767 n
888 3761 n
885 3752 n
885 3746 n
888 3737 n
894 3731 n
903 3728 n
909 3728 n
918 3731 n
924 3737 n
921 3752 m
921 3761 n
918 3767 n
903 3770 m
897 3767 n
891 3761 n
888 3752 n
888 3746 n
891 3737 n
897 3731 n
903 3728 n
912 3758 m
750 4229 m
747 4274 m
741 4268 n
735 4259 n
729 4247 n
726 4232 n
726 4220 n
729 4205 n
735 4193 n
741 4184 n
747 4178 n
741 4268 m
735 4256 n
732 4247 n
729 4232 n
729 4220 n
732 4205 n
735 4196 n
741 4184 n
792 4229 m
771 4235 m
771 4232 n
768 4232 n
768 4235 n
771 4238 n
777 4241 n
789 4241 n
795 4238 n
798 4235 n
801 4229 n
801 4208 n
804 4202 n
807 4199 n
798 4235 m
798 4208 n
801 4202 n
807 4199 n
810 4199 n
798 4229 m
795 4226 n
777 4223 n
768 4220 n
765 4214 n
765 4208 n
768 4202 n
777 4199 n
786 4199 n
792 4202 n
798 4208 n
777 4223 m
771 4220 n
768 4214 n
768 4208 n
771 4202 n
777 4199 n
852 4229 m
825 4274 m
831 4268 n
837 4259 n
843 4247 n
846 4232 n
846 4220 n
843 4205 n
837 4193 n
831 4184 n
825 4178 n
831 4268 m
837 4256 n
840 4247 n
843 4232 n
843 4220 n
840 4205 n
837 4196 n
831 4184 n
750 4229 m
587 3790 m
750 3777 n
912 3758 n
1075 3777 n
1237 3839 n
1400 3918 n
1562 3989 n
1725 4045 n
1887 4089 n
2050 4126 n
2212 4157 n
2375 4185 n
2537 4210 n
2700 4233 n
2862 4254 n
3025 4275 n
3187 4294 n
3350 4312 n
3512 4328 n
3675 4345 n
3837 4360 n
750 3961 m
729 3994 m
729 3931 n
732 3994 m
732 3931 n
732 3964 m
738 3970 n
747 3973 n
753 3973 n
762 3970 n
765 3964 n
765 3931 n
753 3973 m
759 3970 n
762 3964 n
762 3931 n
720 3994 m
732 3994 n
720 3931 m
741 3931 n
753 3931 m
774 3931 n
750 3961 m
/shortdashed f
587 3790 m
750 3961 n
912 4084 n
1075 4173 n
1237 4229 n
1400 4250 n
1562 4247 n
1725 4231 n
1887 4215 n
2050 4201 n
2212 4189 n
2375 4181 n
2537 4175 n
2700 4171 n
2862 4169 n
3025 4167 n
3187 4166 n
3350 4166 n
3512 4167 n
3675 4168 n
3837 4169 n
/solid f
587 3668 m
3837 3668 n
3837 4399 n
587 4399 n
587 3668 n
750 3668 m
750 3718 n
750 4349 m
750 4399 n
912 3668 m
912 3718 n
912 4349 m
912 4399 n
1075 3668 m
1075 3718 n
1075 4349 m
1075 4399 n
1237 3668 m
1237 3718 n
1237 4349 m
1237 4399 n
1400 3668 m
1400 3768 n
1400 4299 m
1400 4399 n
1562 3668 m
1562 3718 n
1562 4349 m
1562 4399 n
1725 3668 m
1725 3718 n
1725 4349 m
1725 4399 n
1887 3668 m
1887 3718 n
1887 4349 m
1887 4399 n
2050 3668 m
2050 3718 n
2050 4349 m
2050 4399 n
2212 3668 m
2212 3768 n
2212 4299 m
2212 4399 n
2375 3668 m
2375 3718 n
2375 4349 m
2375 4399 n
2537 3668 m
2537 3718 n
2537 4349 m
2537 4399 n
2700 3668 m
2700 3718 n
2700 4349 m
2700 4399 n
2862 3668 m
2862 3718 n
2862 4349 m
2862 4399 n
3025 3668 m
3025 3768 n
3025 4299 m
3025 4399 n
3187 3668 m
3187 3718 n
3187 4349 m
3187 4399 n
3350 3668 m
3350 3718 n
3350 4349 m
3350 4399 n
3512 3668 m
3512 3718 n
3512 4349 m
3512 4399 n
3675 3668 m
3675 3718 n
3675 4349 m
3675 4399 n
587 3692 m
637 3692 n
3787 3692 m
3837 3692 n
587 3741 m
637 3741 n
3787 3741 m
3837 3741 n
587 3790 m
687 3790 n
3737 3790 m
3837 3790 n
587 3839 m
637 3839 n
3787 3839 m
3837 3839 n
587 3887 m
637 3887 n
3787 3887 m
3837 3887 n
587 3936 m
637 3936 n
3787 3936 m
3837 3936 n
587 3985 m
637 3985 n
3787 3985 m
3837 3985 n
587 4034 m
687 4034 n
3737 4034 m
3837 4034 n
587 4082 m
637 4082 n
3787 4082 m
3837 4082 n
587 4131 m
637 4131 n
3787 4131 m
3837 4131 n
587 4180 m
637 4180 n
3787 4180 m
3837 4180 n
587 4229 m
637 4229 n
3787 4229 m
3837 4229 n
587 4277 m
687 4277 n
3737 4277 m
3837 4277 n
587 4326 m
637 4326 n
3787 4326 m
3837 4326 n
587 4375 m
637 4375 n
3787 4375 m
3837 4375 n
513 3823 m
504 3820 n
498 3811 n
495 3796 n
495 3787 n
498 3772 n
504 3763 n
513 3760 n
519 3760 n
528 3763 n
534 3772 n
537 3787 n
537 3796 n
534 3811 n
528 3820 n
519 3823 n
513 3823 n
513 3823 m
507 3820 n
504 3817 n
501 3811 n
498 3796 n
498 3787 n
501 3772 n
504 3766 n
507 3763 n
513 3760 n
519 3760 m
525 3763 n
528 3766 n
531 3772 n
534 3787 n
534 3796 n
531 3811 n
528 3817 n
525 3820 n
519 3823 n
429 4067 m
420 4064 n
414 4055 n
411 4040 n
411 4031 n
414 4016 n
420 4007 n
429 4004 n
435 4004 n
444 4007 n
450 4016 n
453 4031 n
453 4040 n
450 4055 n
444 4064 n
435 4067 n
429 4067 n
429 4067 m
423 4064 n
420 4061 n
417 4055 n
414 4040 n
414 4031 n
417 4016 n
420 4010 n
423 4007 n
429 4004 n
435 4004 m
441 4007 n
444 4010 n
447 4016 n
450 4031 n
450 4040 n
447 4055 n
444 4061 n
441 4064 n
435 4067 n
498 4034 m
477 4010 m
474 4007 n
477 4004 n
480 4007 n
477 4010 n
528 4034 m
510 4055 m
516 4058 n
525 4067 n
525 4004 n
522 4064 m
522 4004 n
510 4004 m
537 4004 n
423 4310 m
414 4307 n
408 4298 n
405 4283 n
405 4274 n
408 4259 n
414 4250 n
423 4247 n
429 4247 n
438 4250 n
444 4259 n
447 4274 n
447 4283 n
444 4298 n
438 4307 n
429 4310 n
423 4310 n
423 4310 m
417 4307 n
414 4304 n
411 4298 n
408 4283 n
408 4274 n
411 4259 n
414 4253 n
417 4250 n
423 4247 n
429 4247 m
435 4250 n
438 4253 n
441 4259 n
444 4274 n
444 4283 n
441 4298 n
438 4304 n
435 4307 n
429 4310 n
492 4277 m
471 4253 m
468 4250 n
471 4247 n
474 4250 n
471 4253 n
522 4277 m
498 4298 m
501 4295 n
498 4292 n
495 4295 n
495 4298 n
498 4304 n
501 4307 n
510 4310 n
522 4310 n
531 4307 n
534 4304 n
537 4298 n
537 4292 n
534 4286 n
525 4280 n
510 4274 n
504 4271 n
498 4265 n
495 4256 n
495 4247 n
522 4310 m
528 4307 n
531 4304 n
534 4298 n
534 4292 n
531 4286 n
522 4280 n
510 4274 n
495 4253 m
498 4256 n
504 4256 n
519 4250 n
528 4250 n
534 4253 n
537 4256 n
504 4256 m
519 4247 n
531 4247 n
534 4250 n
537 4256 n
537 4262 n
233 4000 m
302 3973 n
233 4000 m
302 4025 n
243 4000 m
302 4022 n
298 3977 m
298 4022 n
302 3973 m
302 4025 n
269 4072 m
233 4048 m
302 4048 n
233 4052 m
302 4052 n
253 4072 m
278 4072 n
233 4039 m
233 4091 n
253 4091 n
233 4088 n
266 4052 m
266 4072 n
302 4039 m
302 4091 n
282 4091 n
302 4088 n
0 0 4096 4096 s
/solid f
/solid f
587 2937 m
587 2937 m
/solid f
587 2937 p
587 2937 p
587 2937 p
587 2937 p
587 2937 p
587 2937 p
587 2937 p
587 2937 p
587 2937 p
587 2937 p
587 2937 p
587 2937 p
587 2937 p
587 2937 p
587 2937 p
587 2937 p
587 2937 p
587 2937 p
587 2937 p
587 2937 p
587 2937 p
587 2937 p
587 2937 p
587 2937 p
587 2937 p
587 2937 p
587 2937 p
587 2937 p
587 2937 p
587 2937 p
587 2937 p
587 2937 p
587 2937 p
587 2937 p
587 2937 p
587 2937 p
587 2937 p
587 2937 p
587 2937 p
587 2937 p
587 2937 p
587 2937 p
587 2937 p
587 2937 p
587 2937 p
/solid f
587 2937 m
587 2937 m
750 3079 m
726 3073 m
762 3073 n
762 3079 n
759 3085 n
756 3088 n
750 3091 n
741 3091 n
732 3088 n
726 3082 n
723 3073 n
723 3067 n
726 3058 n
732 3052 n
741 3049 n
747 3049 n
756 3052 n
762 3058 n
759 3073 m
759 3082 n
756 3088 n
741 3091 m
735 3088 n
729 3082 n
726 3073 n
726 3067 n
729 3058 n
735 3052 n
741 3049 n
750 3079 m
750 3493 m
747 3538 m
741 3532 n
735 3523 n
729 3511 n
726 3496 n
726 3484 n
729 3469 n
735 3457 n
741 3448 n
747 3442 n
741 3532 m
735 3520 n
732 3511 n
729 3496 n
729 3484 n
732 3469 n
735 3460 n
741 3448 n
792 3493 m
771 3526 m
771 3463 n
774 3526 m
774 3463 n
774 3496 m
780 3502 n
786 3505 n
792 3505 n
801 3502 n
807 3496 n
810 3487 n
810 3481 n
807 3472 n
801 3466 n
792 3463 n
786 3463 n
780 3466 n
774 3472 n
792 3505 m
798 3502 n
804 3496 n
807 3487 n
807 3481 n
804 3472 n
798 3466 n
792 3463 n
762 3526 m
774 3526 n
855 3493 m
828 3538 m
834 3532 n
840 3523 n
846 3511 n
849 3496 n
849 3484 n
846 3469 n
840 3457 n
834 3448 n
828 3442 n
834 3532 m
840 3520 n
843 3511 n
846 3496 n
846 3484 n
843 3469 n
840 3460 n
834 3448 n
750 3493 m
587 2937 m
668 3031 n
750 3079 n
831 3101 n
912 3112 n
993 3122 n
1075 3138 n
1156 3160 n
1237 3189 n
1400 3251 n
1562 3305 n
1725 3347 n
1887 3383 n
2050 3415 n
2212 3443 n
2375 3470 n
2537 3495 n
2700 3520 n
2862 3542 n
3025 3564 n
3187 3585 n
3350 3605 n
3512 3623 n
3675 3641 n
3837 3659 n
750 3228 m
729 3261 m
729 3198 n
732 3261 m
732 3198 n
732 3231 m
738 3237 n
747 3240 n
753 3240 n
762 3237 n
765 3231 n
765 3198 n
753 3240 m
759 3237 n
762 3231 n
762 3198 n
720 3261 m
732 3261 n
720 3198 m
741 3198 n
753 3198 m
774 3198 n
750 3228 m
/shortdashed f
587 2937 m
750 3228 n
912 3408 n
1075 3513 n
1237 3555 n
1400 3541 n
1562 3496 n
1725 3447 n
1887 3409 n
2050 3384 n
2212 3369 n
2375 3361 n
2537 3358 n
2700 3359 n
2862 3363 n
3025 3368 n
3187 3374 n
3350 3381 n
3512 3389 n
3675 3397 n
3837 3405 n
/solid f
587 2937 m
3837 2937 n
3837 3668 n
587 3668 n
587 2937 n
750 2937 m
750 2987 n
750 3618 m
750 3668 n
912 2937 m
912 2987 n
912 3618 m
912 3668 n
1075 2937 m
1075 2987 n
1075 3618 m
1075 3668 n
1237 2937 m
1237 2987 n
1237 3618 m
1237 3668 n
1400 2937 m
1400 3037 n
1400 3568 m
1400 3668 n
1562 2937 m
1562 2987 n
1562 3618 m
1562 3668 n
1725 2937 m
1725 2987 n
1725 3618 m
1725 3668 n
1887 2937 m
1887 2987 n
1887 3618 m
1887 3668 n
2050 2937 m
2050 2987 n
2050 3618 m
2050 3668 n
2212 2937 m
2212 3037 n
2212 3568 m
2212 3668 n
2375 2937 m
2375 2987 n
2375 3618 m
2375 3668 n
2537 2937 m
2537 2987 n
2537 3618 m
2537 3668 n
2700 2937 m
2700 2987 n
2700 3618 m
2700 3668 n
2862 2937 m
2862 2987 n
2862 3618 m
2862 3668 n
3025 2937 m
3025 3037 n
3025 3568 m
3025 3668 n
3187 2937 m
3187 2987 n
3187 3618 m
3187 3668 n
3350 2937 m
3350 2987 n
3350 3618 m
3350 3668 n
3512 2937 m
3512 2987 n
3512 3618 m
3512 3668 n
3675 2937 m
3675 2987 n
3675 3618 m
3675 3668 n
587 2996 m
637 2996 n
3787 2996 m
3837 2996 n
587 3054 m
637 3054 n
3787 3054 m
3837 3054 n
587 3113 m
637 3113 n
3787 3113 m
3837 3113 n
587 3171 m
637 3171 n
3787 3171 m
3837 3171 n
587 3230 m
687 3230 n
3737 3230 m
3837 3230 n
587 3288 m
637 3288 n
3787 3288 m
3837 3288 n
587 3347 m
637 3347 n
3787 3347 m
3837 3347 n
587 3405 m
637 3405 n
3787 3405 m
3837 3405 n
587 3464 m
637 3464 n
3787 3464 m
3837 3464 n
587 3522 m
687 3522 n
3737 3522 m
3837 3522 n
587 3581 m
637 3581 n
3787 3581 m
3837 3581 n
587 3639 m
637 3639 n
3787 3639 m
3837 3639 n
584 2850 m
575 2847 n
569 2838 n
566 2823 n
566 2814 n
569 2799 n
575 2790 n
584 2787 n
590 2787 n
599 2790 n
605 2799 n
608 2814 n
608 2823 n
605 2838 n
599 2847 n
590 2850 n
584 2850 n
584 2850 m
578 2847 n
575 2844 n
572 2838 n
569 2823 n
569 2814 n
572 2799 n
575 2793 n
578 2790 n
584 2787 n
590 2787 m
596 2790 n
599 2793 n
602 2799 n
605 2814 n
605 2823 n
602 2838 n
599 2844 n
596 2847 n
590 2850 n
1385 2850 m
1379 2820 n
1379 2820 m
1385 2826 n
1394 2829 n
1403 2829 n
1412 2826 n
1418 2820 n
1421 2811 n
1421 2805 n
1418 2796 n
1412 2790 n
1403 2787 n
1394 2787 n
1385 2790 n
1382 2793 n
1379 2799 n
1379 2802 n
1382 2805 n
1385 2802 n
1382 2799 n
1403 2829 m
1409 2826 n
1415 2820 n
1418 2811 n
1418 2805 n
1415 2796 n
1409 2790 n
1403 2787 n
1385 2850 m
1415 2850 n
1385 2847 m
1400 2847 n
1415 2850 n
2165 2838 m
2171 2841 n
2180 2850 n
2180 2787 n
2177 2847 m
2177 2787 n
2165 2787 m
2192 2787 n
2243 2817 m
2234 2850 m
2225 2847 n
2219 2838 n
2216 2823 n
2216 2814 n
2219 2799 n
2225 2790 n
2234 2787 n
2240 2787 n
2249 2790 n
2255 2799 n
2258 2814 n
2258 2823 n
2255 2838 n
2249 2847 n
2240 2850 n
2234 2850 n
2234 2850 m
2228 2847 n
2225 2844 n
2222 2838 n
2219 2823 n
2219 2814 n
2222 2799 n
2225 2793 n
2228 2790 n
2234 2787 n
2240 2787 m
2246 2790 n
2249 2793 n
2252 2799 n
2255 2814 n
2255 2823 n
2252 2838 n
2249 2844 n
2246 2847 n
2240 2850 n
2978 2838 m
2984 2841 n
2993 2850 n
2993 2787 n
2990 2847 m
2990 2787 n
2978 2787 m
3005 2787 n
3056 2817 m
3035 2850 m
3029 2820 n
3029 2820 m
3035 2826 n
3044 2829 n
3053 2829 n
3062 2826 n
3068 2820 n
3071 2811 n
3071 2805 n
3068 2796 n
3062 2790 n
3053 2787 n
3044 2787 n
3035 2790 n
3032 2793 n
3029 2799 n
3029 2802 n
3032 2805 n
3035 2802 n
3032 2799 n
3053 2829 m
3059 2826 n
3065 2820 n
3068 2811 n
3068 2805 n
3065 2796 n
3059 2790 n
3053 2787 n
3035 2850 m
3065 2850 n
3035 2847 m
3050 2847 n
3065 2850 n
3789 2838 m
3792 2835 n
3789 2832 n
3786 2835 n
3786 2838 n
3789 2844 n
3792 2847 n
3801 2850 n
3813 2850 n
3822 2847 n
3825 2844 n
3828 2838 n
3828 2832 n
3825 2826 n
3816 2820 n
3801 2814 n
3795 2811 n
3789 2805 n
3786 2796 n
3786 2787 n
3813 2850 m
3819 2847 n
3822 2844 n
3825 2838 n
3825 2832 n
3822 2826 n
3813 2820 n
3801 2814 n
3786 2793 m
3789 2796 n
3795 2796 n
3810 2790 n
3819 2790 n
3825 2793 n
3828 2796 n
3795 2796 m
3810 2787 n
3822 2787 n
3825 2790 n
3828 2796 n
3828 2802 n
3873 2817 m
3864 2850 m
3855 2847 n
3849 2838 n
3846 2823 n
3846 2814 n
3849 2799 n
3855 2790 n
3864 2787 n
3870 2787 n
3879 2790 n
3885 2799 n
3888 2814 n
3888 2823 n
3885 2838 n
3879 2847 n
3870 2850 n
3864 2850 n
3864 2850 m
3858 2847 n
3855 2844 n
3852 2838 n
3849 2823 n
3849 2814 n
3852 2799 n
3855 2793 n
3858 2790 n
3864 2787 n
3870 2787 m
3876 2790 n
3879 2793 n
3882 2799 n
3885 2814 n
3885 2823 n
3882 2838 n
3879 2844 n
3876 2847 n
3870 2850 n
513 2970 m
504 2967 n
498 2958 n
495 2943 n
495 2934 n
498 2919 n
504 2910 n
513 2907 n
519 2907 n
528 2910 n
534 2919 n
537 2934 n
537 2943 n
534 2958 n
528 2967 n
519 2970 n
513 2970 n
513 2970 m
507 2967 n
504 2964 n
501 2958 n
498 2943 n
498 2934 n
501 2919 n
504 2913 n
507 2910 n
513 2907 n
519 2907 m
525 2910 n
528 2913 n
531 2919 n
534 2934 n
534 2943 n
531 2958 n
528 2964 n
525 2967 n
519 2970 n
429 3263 m
420 3260 n
414 3251 n
411 3236 n
411 3227 n
414 3212 n
420 3203 n
429 3200 n
435 3200 n
444 3203 n
450 3212 n
453 3227 n
453 3236 n
450 3251 n
444 3260 n
435 3263 n
429 3263 n
429 3263 m
423 3260 n
420 3257 n
417 3251 n
414 3236 n
414 3227 n
417 3212 n
420 3206 n
423 3203 n
429 3200 n
435 3200 m
441 3203 n
444 3206 n
447 3212 n
450 3227 n
450 3236 n
447 3251 n
444 3257 n
441 3260 n
435 3263 n
498 3230 m
477 3206 m
474 3203 n
477 3200 n
480 3203 n
477 3206 n
528 3230 m
510 3251 m
516 3254 n
525 3263 n
525 3200 n
522 3260 m
522 3200 n
510 3200 m
537 3200 n
423 3555 m
414 3552 n
408 3543 n
405 3528 n
405 3519 n
408 3504 n
414 3495 n
423 3492 n
429 3492 n
438 3495 n
444 3504 n
447 3519 n
447 3528 n
444 3543 n
438 3552 n
429 3555 n
423 3555 n
423 3555 m
417 3552 n
414 3549 n
411 3543 n
408 3528 n
408 3519 n
411 3504 n
414 3498 n
417 3495 n
423 3492 n
429 3492 m
435 3495 n
438 3498 n
441 3504 n
444 3519 n
444 3528 n
441 3543 n
438 3549 n
435 3552 n
429 3555 n
492 3522 m
471 3498 m
468 3495 n
471 3492 n
474 3495 n
471 3498 n
522 3522 m
498 3543 m
501 3540 n
498 3537 n
495 3540 n
495 3543 n
498 3549 n
501 3552 n
510 3555 n
522 3555 n
531 3552 n
534 3549 n
537 3543 n
537 3537 n
534 3531 n
525 3525 n
510 3519 n
504 3516 n
498 3510 n
495 3501 n
495 3492 n
522 3555 m
528 3552 n
531 3549 n
534 3543 n
534 3537 n
531 3531 n
522 3525 n
510 3519 n
495 3498 m
498 3501 n
504 3501 n
519 3495 n
528 3495 n
534 3498 n
537 3501 n
504 3501 m
519 3492 n
531 3492 n
534 3495 n
537 3501 n
537 3507 n
2676 2725 m
2676 2676 n
2680 2666 n
2686 2660 n
2696 2656 n
2702 2656 n
2712 2660 n
2718 2666 n
2722 2676 n
2722 2725 n
2680 2725 m
2680 2676 n
2683 2666 n
2690 2660 n
2696 2656 n
2666 2725 m
2690 2725 n
2712 2725 m
2732 2725 n
233 3269 m
302 3242 n
233 3269 m
302 3294 n
243 3269 m
302 3291 n
298 3246 m
298 3291 n
302 3242 m
302 3294 n
269 3341 m
233 3317 m
302 3317 n
233 3321 m
302 3321 n
253 3341 m
278 3341 n
233 3308 m
233 3360 n
253 3360 n
233 3357 n
266 3321 m
266 3341 n
302 3308 m
302 3360 n
282 3360 n
302 3357 n
0 0 4096 4096 s
/solid f
/solid f
587 1698 m
587 1698 m
/solid f
587 1698 p
587 1698 p
587 1698 p
587 1698 p
587 1698 p
587 1698 p
587 1698 p
587 1698 p
587 1698 p
587 1698 p
587 1698 p
587 1698 p
587 1698 p
587 1698 p
587 1698 p
587 1698 p
587 1698 p
587 1698 p
587 1698 p
587 1698 p
587 1698 p
587 1698 p
587 1698 p
587 1698 p
587 1698 p
587 1698 p
587 1698 p
587 1698 p
587 1698 p
587 1698 p
587 1698 p
587 1698 p
587 1698 p
587 1698 p
587 1698 p
587 1698 p
587 1698 p
587 1698 p
587 1698 p
587 1698 p
587 1698 p
587 1698 p
587 1698 p
587 1698 p
587 1698 p
587 1698 p
587 1698 p
587 1698 p
587 1698 p
587 1698 p
587 1698 p
587 1698 p
587 1698 p
587 1698 p
587 1698 p
587 1698 p
587 1698 p
587 1698 p
587 1698 p
587 1698 p
587 1698 p
587 1698 p
587 1698 p
587 1698 p
587 1698 p
587 1698 p
587 1698 p
587 1698 p
587 1698 p
587 1698 p
587 1698 p
587 1698 p
587 1698 p
587 1698 p
587 1698 p
587 1698 p
587 1698 p
587 1698 p
587 1698 p
587 1698 p
587 1698 p
587 1698 p
587 1698 p
587 1698 p
587 1698 p
587 1698 p
587 1698 p
587 1698 p
587 1698 p
587 1698 p
587 1698 p
587 1698 p
587 1698 p
587 1698 p
587 1698 p
/solid f
587 1698 m
587 1698 m
1562 1776 m
1538 1770 m
1574 1770 n
1574 1776 n
1571 1782 n
1568 1785 n
1562 1788 n
1553 1788 n
1544 1785 n
1538 1779 n
1535 1770 n
1535 1764 n
1538 1755 n
1544 1749 n
1553 1746 n
1559 1746 n
1568 1749 n
1574 1755 n
1571 1770 m
1571 1779 n
1568 1785 n
1553 1788 m
1547 1785 n
1541 1779 n
1538 1770 n
1538 1764 n
1541 1755 n
1547 1749 n
1553 1746 n
1562 1776 m
587 1284 m
750 1410 n
912 1497 n
1075 1575 n
1237 1655 n
1400 1726 n
1562 1776 n
1725 1805 n
1887 1831 n
2050 1842 n
2212 1845 n
2375 1843 n
2537 1837 n
2700 1828 n
2862 1817 n
3025 1805 n
3187 1791 n
3350 1777 n
3512 1761 n
3675 1746 n
3837 1729 n
1562 1572 m
1541 1605 m
1541 1542 n
1544 1605 m
1544 1542 n
1544 1575 m
1550 1581 n
1559 1584 n
1565 1584 n
1574 1581 n
1577 1575 n
1577 1542 n
1565 1584 m
1571 1581 n
1574 1575 n
1574 1542 n
1532 1605 m
1544 1605 n
1532 1542 m
1553 1542 n
1565 1542 m
1586 1542 n
1562 1572 m
/shortdashed f
587 1216 m
750 1391 n
912 1502 n
1075 1569 n
1237 1600 n
1400 1598 n
1562 1572 n
1725 1533 n
1887 1492 n
2050 1471 n
2212 1470 n
2375 1481 n
2537 1488 n
2700 1496 n
2862 1504 n
3025 1513 n
3187 1522 n
3350 1531 n
3512 1539 n
3675 1548 n
3837 1556 n
/solid f
587 1150 m
3837 1150 n
3837 1881 n
587 1881 n
587 1150 n
750 1150 m
750 1200 n
750 1831 m
750 1881 n
912 1150 m
912 1200 n
912 1831 m
912 1881 n
1075 1150 m
1075 1200 n
1075 1831 m
1075 1881 n
1237 1150 m
1237 1200 n
1237 1831 m
1237 1881 n
1400 1150 m
1400 1250 n
1400 1781 m
1400 1881 n
1562 1150 m
1562 1200 n
1562 1831 m
1562 1881 n
1725 1150 m
1725 1200 n
1725 1831 m
1725 1881 n
1887 1150 m
1887 1200 n
1887 1831 m
1887 1881 n
2050 1150 m
2050 1200 n
2050 1831 m
2050 1881 n
2212 1150 m
2212 1250 n
2212 1781 m
2212 1881 n
2375 1150 m
2375 1200 n
2375 1831 m
2375 1881 n
2537 1150 m
2537 1200 n
2537 1831 m
2537 1881 n
2700 1150 m
2700 1200 n
2700 1831 m
2700 1881 n
2862 1150 m
2862 1200 n
2862 1831 m
2862 1881 n
3025 1150 m
3025 1250 n
3025 1781 m
3025 1881 n
3187 1150 m
3187 1200 n
3187 1831 m
3187 1881 n
3350 1150 m
3350 1200 n
3350 1831 m
3350 1881 n
3512 1150 m
3512 1200 n
3512 1831 m
3512 1881 n
3675 1150 m
3675 1200 n
3675 1831 m
3675 1881 n
587 1187 m
637 1187 n
3787 1187 m
3837 1187 n
587 1260 m
637 1260 n
3787 1260 m
3837 1260 n
587 1333 m
687 1333 n
3737 1333 m
3837 1333 n
587 1406 m
637 1406 n
3787 1406 m
3837 1406 n
587 1479 m
637 1479 n
3787 1479 m
3837 1479 n
587 1552 m
637 1552 n
3787 1552 m
3837 1552 n
587 1625 m
637 1625 n
3787 1625 m
3837 1625 n
587 1698 m
687 1698 n
3737 1698 m
3837 1698 n
587 1772 m
637 1772 n
3787 1772 m
3837 1772 n
587 1845 m
637 1845 n
3787 1845 m
3837 1845 n
1526 516 m
1529 507 n
1529 525 n
1526 516 n
1520 522 n
1511 525 n
1505 525 n
1496 522 n
1490 516 n
1487 510 n
1484 501 n
1484 486 n
1487 477 n
1490 471 n
1496 465 n
1505 462 n
1511 462 n
1520 465 n
1526 471 n
1505 525 m
1499 522 n
1493 516 n
1490 510 n
1487 501 n
1487 486 n
1490 477 n
1493 471 n
1499 465 n
1505 462 n
1526 486 m
1526 462 n
1529 486 m
1529 462 n
1517 486 m
1538 486 n
1580 492 m
1559 498 m
1559 495 n
1556 495 n
1556 498 n
1559 501 n
1565 504 n
1577 504 n
1583 501 n
1586 498 n
1589 492 n
1589 471 n
1592 465 n
1595 462 n
1586 498 m
1586 471 n
1589 465 n
1595 462 n
1598 462 n
1586 492 m
1583 489 n
1565 486 n
1556 483 n
1553 477 n
1553 471 n
1556 465 n
1565 462 n
1574 462 n
1580 465 n
1586 471 n
1565 486 m
1559 483 n
1556 477 n
1556 471 n
1559 465 n
1565 462 n
1640 492 m
1619 504 m
1619 462 n
1622 504 m
1622 462 n
1622 495 m
1628 501 n
1637 504 n
1643 504 n
1652 501 n
1655 495 n
1655 462 n
1643 504 m
1649 501 n
1652 495 n
1652 462 n
1655 495 m
1661 501 n
1670 504 n
1676 504 n
1685 501 n
1688 495 n
1688 462 n
1676 504 m
1682 501 n
1685 495 n
1685 462 n
1610 504 m
1622 504 n
1610 462 m
1631 462 n
1643 462 m
1664 462 n
1676 462 m
1697 462 n
1739 492 m
1718 504 m
1718 462 n
1721 504 m
1721 462 n
1721 495 m
1727 501 n
1736 504 n
1742 504 n
1751 501 n
1754 495 n
1754 462 n
1742 504 m
1748 501 n
1751 495 n
1751 462 n
1754 495 m
1760 501 n
1769 504 n
1775 504 n
1784 501 n
1787 495 n
1787 462 n
1775 504 m
1781 501 n
1784 495 n
1784 462 n
1709 504 m
1721 504 n
1709 462 m
1730 462 n
1742 462 m
1763 462 n
1775 462 m
1796 462 n
1838 492 m
1814 486 m
1850 486 n
1850 492 n
1847 498 n
1844 501 n
1838 504 n
1829 504 n
1820 501 n
1814 495 n
1811 486 n
1811 480 n
1814 471 n
1820 465 n
1829 462 n
1835 462 n
1844 465 n
1850 471 n
1847 486 m
1847 495 n
1844 501 n
1829 504 m
1823 501 n
1817 495 n
1814 486 n
1814 480 n
1817 471 n
1823 465 n
1829 462 n
1895 492 m
1874 525 m
1874 462 n
1877 525 m
1877 462 n
1865 525 m
1877 525 n
1865 462 m
1886 462 n
1928 492 m
1976 492 m
1952 486 m
1988 486 n
1988 492 n
1985 498 n
1982 501 n
1976 504 n
1967 504 n
1958 501 n
1952 495 n
1949 486 n
1949 480 n
1952 471 n
1958 465 n
1967 462 n
1973 462 n
1982 465 n
1988 471 n
1985 486 m
1985 495 n
1982 501 n
1967 504 m
1961 501 n
1955 495 n
1952 486 n
1952 480 n
1955 471 n
1961 465 n
1967 462 n
2033 492 m
2012 525 m
2012 474 n
2015 465 n
2021 462 n
2027 462 n
2033 465 n
2036 471 n
2015 525 m
2015 474 n
2018 465 n
2021 462 n
2003 504 m
2027 504 n
2078 492 m
2126 492 m
2105 498 m
2105 495 n
2102 495 n
2102 498 n
2105 501 n
2111 504 n
2123 504 n
2129 501 n
2132 498 n
2135 492 n
2135 471 n
2138 465 n
2141 462 n
2132 498 m
2132 471 n
2135 465 n
2141 462 n
2144 462 n
2132 492 m
2129 489 n
2111 486 n
2102 483 n
2099 477 n
2099 471 n
2102 465 n
2111 462 n
2120 462 n
2126 465 n
2132 471 n
2111 486 m
2105 483 n
2102 477 n
2102 471 n
2105 465 n
2111 462 n
2186 492 m
2165 525 m
2165 462 n
2168 525 m
2168 462 n
2156 525 m
2168 525 n
2156 462 m
2177 462 n
2219 492 m
2198 462 m
2195 465 n
2198 468 n
2201 465 n
2201 459 n
2198 453 n
2195 450 n
2249 492 m
2297 492 m
2276 525 m
2276 462 n
2279 525 m
2279 462 n
2297 507 m
2297 483 n
2267 525 m
2315 525 n
2315 507 n
2312 525 n
2279 495 m
2297 495 n
2267 462 m
2288 462 n
2357 492 m
2336 525 m
2333 522 n
2336 519 n
2339 522 n
2336 525 n
2336 504 m
2336 462 n
2339 504 m
2339 462 n
2327 504 m
2339 504 n
2327 462 m
2348 462 n
2390 492 m
2378 504 m
2372 501 n
2369 498 n
2366 492 n
2366 486 n
2369 480 n
2372 477 n
2378 474 n
2384 474 n
2390 477 n
2393 480 n
2396 486 n
2396 492 n
2393 498 n
2390 501 n
2384 504 n
2378 504 n
2372 501 m
2369 495 n
2369 483 n
2372 477 n
2390 477 m
2393 483 n
2393 495 n
2390 501 n
2393 498 m
2396 501 n
2402 504 n
2402 501 n
2396 501 n
2369 480 m
2366 477 n
2363 471 n
2363 468 n
2366 462 n
2375 459 n
2390 459 n
2399 456 n
2402 453 n
2363 468 m
2366 465 n
2375 462 n
2390 462 n
2399 459 n
2402 453 n
2402 450 n
2399 444 n
2390 441 n
2372 441 n
2363 444 n
2360 450 n
2360 453 n
2363 459 n
2372 462 n
2447 492 m
2426 468 m
2423 465 n
2426 462 n
2429 465 n
2426 468 n
2477 492 m
2480 498 m
2483 504 n
2483 492 n
2480 498 n
2477 501 n
2471 504 n
2459 504 n
2453 501 n
2450 498 n
2450 492 n
2453 489 n
2459 486 n
2474 480 n
2480 477 n
2483 474 n
2450 495 m
2453 492 n
2459 489 n
2474 483 n
2480 480 n
2483 477 n
2483 468 n
2480 465 n
2474 462 n
2462 462 n
2456 465 n
2453 468 n
2450 474 n
2450 462 n
2453 468 n
2528 492 m
2576 492 m
2552 513 m
2555 510 n
2552 507 n
2549 510 n
2549 513 n
2552 519 n
2555 522 n
2564 525 n
2576 525 n
2585 522 n
2588 516 n
2588 507 n
2585 501 n
2576 498 n
2567 498 n
2576 525 m
2582 522 n
2585 516 n
2585 507 n
2582 501 n
2576 498 n
2576 498 m
2582 495 n
2588 489 n
2591 483 n
2591 474 n
2588 468 n
2585 465 n
2576 462 n
2564 462 n
2555 465 n
2552 468 n
2549 474 n
2549 477 n
2552 480 n
2555 477 n
2552 474 n
2585 492 m
2588 483 n
2588 474 n
2585 468 n
2582 465 n
2576 462 n
2636 492 m
2684 492 m
2663 498 m
2663 495 n
2660 495 n
2660 498 n
2663 501 n
2669 504 n
2681 504 n
2687 501 n
2690 498 n
2693 492 n
2693 471 n
2696 465 n
2699 462 n
2690 498 m
2690 471 n
2693 465 n
2699 462 n
2702 462 n
2690 492 m
2687 489 n
2669 486 n
2660 483 n
2657 477 n
2657 471 n
2660 465 n
2669 462 n
2678 462 n
2684 465 n
2690 471 n
2669 486 m
2663 483 n
2660 477 n
2660 471 n
2663 465 n
2669 462 n
2744 492 m
2723 504 m
2723 462 n
2726 504 m
2726 462 n
2726 495 m
2732 501 n
2741 504 n
2747 504 n
2756 501 n
2759 495 n
2759 462 n
2747 504 m
2753 501 n
2756 495 n
2756 462 n
2714 504 m
2726 504 n
2714 462 m
2735 462 n
2747 462 m
2768 462 n
2810 492 m
2819 525 m
2819 462 n
2822 525 m
2822 462 n
2819 495 m
2813 501 n
2807 504 n
2801 504 n
2792 501 n
2786 495 n
2783 486 n
2783 480 n
2786 471 n
2792 465 n
2801 462 n
2807 462 n
2813 465 n
2819 471 n
2801 504 m
2795 501 n
2789 495 n
2786 486 n
2786 480 n
2789 471 n
2795 465 n
2801 462 n
2810 525 m
2822 525 n
2819 462 m
2831 462 n
2873 492 m
2921 492 m
2921 519 m
2921 462 n
2924 525 m
2924 462 n
2924 525 m
2891 480 n
2939 480 n
2912 462 m
2933 462 n
584 1063 m
575 1060 n
569 1051 n
566 1036 n
566 1027 n
569 1012 n
575 1003 n
584 1000 n
590 1000 n
599 1003 n
605 1012 n
608 1027 n
608 1036 n
605 1051 n
599 1060 n
590 1063 n
584 1063 n
584 1063 m
578 1060 n
575 1057 n
572 1051 n
569 1036 n
569 1027 n
572 1012 n
575 1006 n
578 1003 n
584 1000 n
590 1000 m
596 1003 n
599 1006 n
602 1012 n
605 1027 n
605 1036 n
602 1051 n
599 1057 n
596 1060 n
590 1063 n
1385 1063 m
1379 1033 n
1379 1033 m
1385 1039 n
1394 1042 n
1403 1042 n
1412 1039 n
1418 1033 n
1421 1024 n
1421 1018 n
1418 1009 n
1412 1003 n
1403 1000 n
1394 1000 n
1385 1003 n
1382 1006 n
1379 1012 n
1379 1015 n
1382 1018 n
1385 1015 n
1382 1012 n
1403 1042 m
1409 1039 n
1415 1033 n
1418 1024 n
1418 1018 n
1415 1009 n
1409 1003 n
1403 1000 n
1385 1063 m
1415 1063 n
1385 1060 m
1400 1060 n
1415 1063 n
2165 1051 m
2171 1054 n
2180 1063 n
2180 1000 n
2177 1060 m
2177 1000 n
2165 1000 m
2192 1000 n
2243 1030 m
2234 1063 m
2225 1060 n
2219 1051 n
2216 1036 n
2216 1027 n
2219 1012 n
2225 1003 n
2234 1000 n
2240 1000 n
2249 1003 n
2255 1012 n
2258 1027 n
2258 1036 n
2255 1051 n
2249 1060 n
2240 1063 n
2234 1063 n
2234 1063 m
2228 1060 n
2225 1057 n
2222 1051 n
2219 1036 n
2219 1027 n
2222 1012 n
2225 1006 n
2228 1003 n
2234 1000 n
2240 1000 m
2246 1003 n
2249 1006 n
2252 1012 n
2255 1027 n
2255 1036 n
2252 1051 n
2249 1057 n
2246 1060 n
2240 1063 n
2978 1051 m
2984 1054 n
2993 1063 n
2993 1000 n
2990 1060 m
2990 1000 n
2978 1000 m
3005 1000 n
3056 1030 m
3035 1063 m
3029 1033 n
3029 1033 m
3035 1039 n
3044 1042 n
3053 1042 n
3062 1039 n
3068 1033 n
3071 1024 n
3071 1018 n
3068 1009 n
3062 1003 n
3053 1000 n
3044 1000 n
3035 1003 n
3032 1006 n
3029 1012 n
3029 1015 n
3032 1018 n
3035 1015 n
3032 1012 n
3053 1042 m
3059 1039 n
3065 1033 n
3068 1024 n
3068 1018 n
3065 1009 n
3059 1003 n
3053 1000 n
3035 1063 m
3065 1063 n
3035 1060 m
3050 1060 n
3065 1063 n
3789 1051 m
3792 1048 n
3789 1045 n
3786 1048 n
3786 1051 n
3789 1057 n
3792 1060 n
3801 1063 n
3813 1063 n
3822 1060 n
3825 1057 n
3828 1051 n
3828 1045 n
3825 1039 n
3816 1033 n
3801 1027 n
3795 1024 n
3789 1018 n
3786 1009 n
3786 1000 n
3813 1063 m
3819 1060 n
3822 1057 n
3825 1051 n
3825 1045 n
3822 1039 n
3813 1033 n
3801 1027 n
3786 1006 m
3789 1009 n
3795 1009 n
3810 1003 n
3819 1003 n
3825 1006 n
3828 1009 n
3795 1009 m
3810 1000 n
3822 1000 n
3825 1003 n
3828 1009 n
3828 1015 n
3873 1030 m
3864 1063 m
3855 1060 n
3849 1051 n
3846 1036 n
3846 1027 n
3849 1012 n
3855 1003 n
3864 1000 n
3870 1000 n
3879 1003 n
3885 1012 n
3888 1027 n
3888 1036 n
3885 1051 n
3879 1060 n
3870 1063 n
3864 1063 n
3864 1063 m
3858 1060 n
3855 1057 n
3852 1051 n
3849 1036 n
3849 1027 n
3852 1012 n
3855 1006 n
3858 1003 n
3864 1000 n
3870 1000 m
3876 1003 n
3879 1006 n
3882 1012 n
3885 1027 n
3885 1036 n
3882 1051 n
3879 1057 n
3876 1060 n
3870 1063 n
330 1330 m
384 1330 n
432 1333 m
423 1366 m
414 1363 n
408 1354 n
405 1339 n
405 1330 n
408 1315 n
414 1306 n
423 1303 n
429 1303 n
438 1306 n
444 1315 n
447 1330 n
447 1339 n
444 1354 n
438 1363 n
429 1366 n
423 1366 n
423 1366 m
417 1363 n
414 1360 n
411 1354 n
408 1339 n
408 1330 n
411 1315 n
414 1309 n
417 1306 n
423 1303 n
429 1303 m
435 1306 n
438 1309 n
441 1315 n
444 1330 n
444 1339 n
441 1354 n
438 1360 n
435 1363 n
429 1366 n
492 1333 m
471 1309 m
468 1306 n
471 1303 n
474 1306 n
471 1309 n
522 1333 m
498 1354 m
501 1351 n
498 1348 n
495 1351 n
495 1354 n
498 1360 n
501 1363 n
510 1366 n
522 1366 n
531 1363 n
534 1360 n
537 1354 n
537 1348 n
534 1342 n
525 1336 n
510 1330 n
504 1327 n
498 1321 n
495 1312 n
495 1303 n
522 1366 m
528 1363 n
531 1360 n
534 1354 n
534 1348 n
531 1342 n
522 1336 n
510 1330 n
495 1309 m
498 1312 n
504 1312 n
519 1306 n
528 1306 n
534 1309 n
537 1312 n
504 1312 m
519 1303 n
531 1303 n
534 1306 n
537 1312 n
537 1318 n
513 1731 m
504 1728 n
498 1719 n
495 1704 n
495 1695 n
498 1680 n
504 1671 n
513 1668 n
519 1668 n
528 1671 n
534 1680 n
537 1695 n
537 1704 n
534 1719 n
528 1728 n
519 1731 n
513 1731 n
513 1731 m
507 1728 n
504 1725 n
501 1719 n
498 1704 n
498 1695 n
501 1680 n
504 1674 n
507 1671 n
513 1668 n
519 1668 m
525 1671 n
528 1674 n
531 1680 n
534 1695 n
534 1704 n
531 1719 n
528 1725 n
525 1728 n
519 1731 n
2676 938 m
2676 889 n
2680 879 n
2686 873 n
2696 869 n
2702 869 n
2712 873 n
2718 879 n
2722 889 n
2722 938 n
2680 938 m
2680 889 n
2683 879 n
2690 873 n
2696 869 n
2666 938 m
2690 938 n
2712 938 m
2732 938 n
173 1482 m
242 1455 n
173 1482 m
242 1507 n
183 1482 m
242 1504 n
238 1459 m
238 1504 n
242 1455 m
242 1507 n
209 1554 m
173 1530 m
242 1530 n
173 1534 m
242 1534 n
193 1554 m
218 1554 n
173 1521 m
173 1573 n
193 1573 n
173 1570 n
206 1534 m
206 1554 n
242 1521 m
242 1573 n
222 1573 n
242 1570 n
e
EndPSPlot